\documentclass[traditabstract,twocolumns]{aa}

\usepackage{txfonts}
\usepackage{graphicx}	
\usepackage{amsmath}	
\usepackage{amssymb}	
\usepackage{xcolor}     
\usepackage{soul}       
\usepackage{array}
\usepackage[colorlinks=true,linkcolor=blue,citecolor=blue, urlcolor=blue]{hyperref}
\bibpunct{(}{)}{;}{a}{}{,}

\usepackage{fontawesome}  

\definecolor{fredcolor}{RGB}{204, 204, 0}
\definecolor{changesgreyblue}{RGB}{78, 129, 166}

\defcitealias{Joseph2019}{J19}

\newcommand{\customitem}{\item[$\bullet$]}

\newcommand{\lenstro}{\textsc{Lenstronomy}\xspace}
\newcommand{\slit}{\textsc{SLIT}\xspace}
\newcommand{\slitro}{\textsc{SLITronomy}\xspace}
\newcommand{\sersic}{S\'ersic\xspace}

\newcommand{\mat}[1]{\ensuremath{\mathsf{#1}}\xspace}
\newcommand{\source}{\mat{s}}
\newcommand{\lensconv}{\mat{g_H}}
\newcommand{\lens}{\mat{g}}
\newcommand{\data}{\mat{\tilde{y}}}
\newcommand{\model}{\mat{y}}
\newcommand{\lensingop}{\mat{F}}
\newcommand{\convop}{\mat{H}}
\newcommand{\noise}{\mat{n}}
\newcommand{\psfkernel}{\ensuremath{K}\xspace}
\newcommand{\starletsopexplicit}[1]{\ensuremath{\Phi^\top_{#1}}\xspace}
\newcommand{\starletsinvopexplicit}[1]{\ensuremath{\Phi_{#1}}\xspace}
\newcommand{\starletsop}{\ensuremath{\Phi^\top}\xspace}
\newcommand{\starletsinvop}{\ensuremath{\Phi}\xspace}
\newcommand{\Jscales}[1]{\ensuremath{J_{#1}}\xspace}
\newcommand{\Wweights}[1]{\ensuremath{W_{#1}}\xspace}
\newcommand{\prior}[1]{\ensuremath{\mathcal{P}\left(#1\right)}\xspace}
\newcommand{\likelihood}[1]{\ensuremath{\mathcal{L}\left(#1\right)}\xspace}

\newcommand{\pixsizeratio}{\ensuremath{r_{\rm pix}}\xspace}

\newcommand{\normone}[1]{\ensuremath{\left\lVert #1 \right\rVert_1}}
\newcommand{\normtwo}[1]{\ensuremath{\left\lVert #1 \right\rVert_2}}

\newcommand{\zeroneufcinquanteneuf}{SDSS\,J0959$-$0410\xspace}
\newcommand{\douzecinquante}{SDSS\,J1250$+$0523\xspace}
\newcommand{\seizevingtsept}{SDSS\,J1627$-$0053\xspace}
\newcommand{\seizetrente}{SDSS\,J1630$-$4520\xspace}

\DeclareMathOperator*{\argmin}{arg\,min}

\begin{document}


\title{\slitro: towards a fully wavelet-based\\strong lensing inversion technique}

\titlerunning{Wavelet-based modelling with \slitro}

\author{
A. Galan\inst{\ref{epfl}}, A. Peel\inst{\ref{epfl}}, R. Joseph\inst{\ref{princeton}}, F. Courbin\inst{\ref{epfl}}, \and J.-L. Starck\inst{\ref{cea}}
}

\institute{
Institute of Physics, Laboratory of Astrophysics, Ecole Polytechnique 
F\'ed\'erale de Lausanne (EPFL), Observatoire de Sauverny, 1290 Versoix, 
Switzerland \label{epfl}
\goodbreak 
\and
Department of Astrophysical Sciences, Princeton University, Princeton, NJ 08544, USA \label{princeton}
\goodbreak
\and
AIM, CEA, CNRS, Universit\'e Paris-Saclay, Universit\'e de Paris, F-91191 Gif-sur-Yvette, France \label{cea}
}

\abstract{
Strong gravitational lensing provides a wealth of astrophysical information on the baryonic and dark matter content of galaxies. It also serves as a valuable cosmological probe by allowing us to measure the Hubble constant independently of other methods. These applications all require the difficult task of inverting the lens equation and simultaneously reconstructing the mass profile of the lens along with the original light profile of the unlensed source. As there is no reason for either the lens or the source to be simple, we need methods that both invert the lens equation with a large number of degrees of freedom and also enforce a well-controlled regularisation that avoids the appearance of spurious structures. This can be beautifully accomplished by representing signals in wavelet space. Building on the Sparse Lens Inversion Technique (SLIT), in this work we present an improved sparsity-based method that describes lensed sources using wavelets and optimises over the parameters given an analytical lens mass profile. We apply our technique on simulated HST and E-ELT data, as well as on real HST images of lenses from the Sloan Lens ACS (SLACS) sample, assuming a lens model. We show that wavelets allow us to reconstruct lensed sources containing detailed substructures when using both present-day data and very high-resolution images expected from future thirty-meter-class telescopes. In the latter case, wavelets moreover provide a much more tractable solution in terms of quality and computation time compared to using a source model that combines smooth analytical profiles and shapelets. Requiring very little human interaction, our flexible pixel-based technique fits into the ongoing effort to devise automated modelling schemes. It can be incorporated in the standard workflow of sampling analytical lens model parameters while modelling the source on a pixelated grid. The method, which we call \slitro, is freely available as a new plug-in to the modelling software \lenstro.
}


\maketitle

\section{Introduction}

Gravitational lensing offers a window into the nature of many fundamental properties of our universe. The fortuitous alignment of a distant luminous object (i.e., source) and a massive foreground structure (i.e., lens) along the line of sight can produce striking visual distortions that allow us to constrain both astrophysical and cosmological parameters. By magnifying the distant source, strong lensing can reveal detailed structures that would otherwise be unresolvable. Cluster-scale lenses allow us to study the large-scale distribution of mass and the hierarchical evolution of dark matter halos \citep{Natarajan2007,Bhattacharya2013,Han2015}. Galaxy-scale lenses give insights on smaller scales, such as substructure within the main deflector \citep[e.g.][]{Amara2006, Vegetti2010,Hezaveh2016,Birrer2017,Ritondale2019,Gilman2020}, as well as population-level properties of galaxies when large samples can be uniformly selected \citep{Bolton2006,Brownstein2012,Gavazzi2012}. Strongly lensed quasars are particularly useful to constrain black hole evolution over a wide range of redshifts \citep[e.g.][]{Ding2020} and to derive high-precision constraints on the Hubble constant \citep[][now referred as the TDCOSMO collaboration]{Wong2019,Shajib2020,Birrer2020}.


All of the above applications require lens modelling tools. If the underlying cosmology needed to compute distances to the lens and source is not exactly known, models are subject to degeneracies, to modelling assumptions \citep{Xu2016,Sonnenfeld2018,Blum2020} and/or to the inherent geometry and physics of the problem \citep{Saha2000,Schneider2013}. As a result, for any given rescaling of the lens mass, there exists a corresponding rescaling of the lensed source that leads to identical lensing observables apart from the time delays. These degeneracies disappear if the underlying cosmology is known, but even in this case, lens modelling tools must in practice address several key points to work efficiently: (1) reliable reconstruction of the lens mass and light, (2) reliable reconstruction (de-lensing) of the source light, (3) efficient deblending of the lens and source light.

In this work, we focus on the second point, namely source light reconstruction. Various techniques to model the light of the background source have been developed over the years. There are two main families of techniques, which we distinguish as ``analytic'' and ``pixel-based'' methods. While the former assumes a set of analytic functions through which the source is represented, the latter uses pixels directly to describe surface galaxy light distribution. We do not use the term ``free-form techniques'' for pixel-based methods, as a pixelated surface brightness is already a constraining assumption compared to the continuous truth \citep[e.g.][]{TagoreJackson2016}.

There are many public software packages that implement analytic methods for strong lens analysis. For example, \textsc{lenstool} \citep{Kneib2011} and \textsc{glafic} \citep{Oguri2010} provide specialised tools for cluster-scale lenses. For galaxy-galaxy lenses, dedicated software has been developed, such as \textsc{gravlens} \citep{Keeton2011}, \textsc{PixeLens} \citep{Williams2004}, \textsc{AutoLens} \citep{Nightingale2018}, and \lenstro \citep{Birrer2018lenstro}. Typical analytic functions include variations of power-law or cored profiles for the lensing mass, while more elaborate ensembles of basis functions like shapelets \citep{Bernstein2002,Refregier2003,TagoreKeeton2014,Birrer2015} can be used to capture the wider dynamic range of features often exhibited by high-redshift galaxies, especially when they are star-forming. While analytical methods are fast, they have two main disadvantages. First, complex light distributions like galaxies with resolved small-scale features or mergers are poorly modelled. Second, it requires lots of trial and error to infer the initial parameters describing the source shape, hence making it hard to automate these techniques.

Recent developments in deep learning have enabled neural networks, which can be seen as a collection of nested analytic functions of many (sometimes millions of) parameters, to become a viable tool for strong lens modelling. The work of \cite{Hezaveh2017} \citep[and follow-up papers][]{PerreaultLevasseur2017,Morningstar2018,Morningstar2019}, along with \cite{Pearson2019}, have shown promising initial results on simulated data. Coupled with GPUs for training, neural networks can offer vast speed improvements compared to other techniques, although their applicability in practice hinges crucially on the quality of the training data. A key to modern neural network optimisation is automatic differentiation, which also facilitates the creation of fully differentiable pipelines capable of efficiently exploring the large parameter space of strong lens models \citep{Chianese2020}.

Within the family of pixel-based techniques, to which this work belongs, \cite{KochanekNarayan1992} described one of the first methods, which is based on the \textsc{clean} algorithm widely used for processing radio astronomy data \citep{Hogbom1974}. Later, \cite{WarrenDye2003} established the popular semi-linear inversion (SLI) formalism, where the (under-constrained) source surface brightness is optimised through regularised inversion. Its update to include adaptive gridding \citep{DyeWarren2005} solved several issues inherent to the SLI method. \cite{Suyu2006} brought further improvements to the method using the Bayesian formalism to objectively find hyper-parameters that are required for the lens inversion \citep[\textsc{GLEE} modelling code, ][]{SuyuHalkola2010}. Various adaptive pixel grids have been tested to mitigate biases that can arise when inferring lens model parameters while still maintaining a tractable computation time. For instance, the algorithm by \citet{VegettiKoopmans2009} uses a Delaunay tessellation, combined with an iterative scheme, to find deviations from a smooth gravitational potential, and \cite{NightingaleDye2015} uses randomly initialised $h$-means clustering to define the adaptive grid. More recently, \citet{Rizzo2018} and \citet{Powell2020} improved the technique of \citet{VegettiKoopmans2009} with the joint modelling of resolved kinematics and application to interferometry data.

The above-mentioned methods all primarily differ from each other in their constraints on the galaxy surface brightness distribution, independent of the imaging data. These constraints restrict the large parameter space common to pixel-based methods, thus reducing degeneracies and improving convergence of the algorithm. However, they often lack a clear physical interpretation motivated by light distributions of real galaxies. Additionally, they introduce new hyper-parameters, whose optimal values can vary significantly depending on the data and are hence complex to handle.

Drawing from standard concepts in the field of image processing, \cite{Joseph2019} -- hereafter \citetalias{Joseph2019} -- introduced improved assumptions on light distributions through a technique based on sparsity and wavelets called the Sparse Linear Inversion Technique (SLIT), which mitigates both issues mentioned above. The present paper introduces an updated and expanded open-source implementation of \slit, which serves as a plug-in to the modelling software \lenstro. All codes described here are publicly available, and data along with python notebooks used to generate results and figures of this paper are available \href{https://github.com/aymgal/SLITronomy-papers/tree/master/paper_I}{here~\faGithub}.

As a validation of the algorithm, we reconstruct high-resolution source galaxies from of the Sloan Lens Advanced Camera for Surveys (SLACS) sample of strong galaxy-galaxy lenses \citep[][]{Bolton2006}. This dataset has been extensively studied for constraining population-level properties of massive elliptical galaxies. However, little discussion has focused on the quality of source reconstructions, as it is commonly assumed that constraining deflector properties to the percent level can be achieved without high-fidelity source reconstructions. Based on the recent (re)modelling of a subset of the SLACS sample by \cite{Shajib2020slacs}, we compare for the first time source reconstructions from analytical and pixel-based methods on this challenging dataset, consistently in the same modelling environment.

We further apply our modelling method to simulated data that is representative of the quality achievable by future thirty-meter-class telescopes. We show that our method is particularly well-suited to such extremely high resolution imaging data and that modelling the source light profile analytically quickly leads to an intractable problem in practice.

Our paper is organised as follows. We first review some basics of the strong gravitational lensing formalism and its linear approximation in Section \ref{sec:theory}. In Section~\ref{sec:slit} we give an overview of the method and theory behind the original \slit method. We discuss key improvements in Section~\ref{sec:slitronomy} and describe our new code implementation, which we call \slitro. In Section~\ref{sec:hst_data}, we apply our technique to source reconstruction problems: first on simulated data, and then on a subset of SLACS lenses. We consistently compare the results in both cases with reconstructions using shapelets. To demonstrate the feasibility of the approach, we show sampled posteriors over lens model parameters. In Section~\ref{sec:elt_data}, we proceed similarly on E-ELT simulated data and show how sparse reconstruction is suited to such high-resolution data. Finally we summarise and conclude in Section~\ref{sec:conclusion}.

\section{Theoretical aspects of strong gravitational lens inversion \label{sec:theory}}

\subsection{Strong gravitational lensing}

Gravitational lensing arises when a massive foreground object lies on or near the line-of-sight connecting an observer and a distant luminous source. The lensing effect is considered \textit{strong} when the lens galaxy (or cluster of galaxies) is so massive that it produces high order deformations, possibly leading to several images, of the background source\footnote{Image multiplicity is sometimes viewed as a necessary condition for the lensing effect to be considered strong.}. In this case, the source light rays are deflected from their original paths, and the lens equation (see below) displays several solutions.

In the following we write the redshift of the lensed source as $z_{\rm s}$ and the redshift of the lensing galaxy, also called the deflector, as $z_{\rm d}$. We introduce the lensing potential $\psi(\vec{\theta})$, obtained by rescaling and projecting the three-dimensional gravitational potential on the lens plane, then evaluated at angular position \vec{\theta} on the coordinate grid. Light rays from the source are deflected by the deflection angle field $\vec{\alpha}$, which is directly related to the projected potential through its first derivatives: $\vec{\alpha} = \nabla\psi$. The lens equation gives the angular coordinates of deflected light rays from a source at position \vec{\beta} for every position \vec{\theta}:
\begin{align}
    \label{eq:lens_eq}
    \vec{\beta} = \vec{\theta} - \vec{\alpha}(\vec{\theta})\ .
\end{align}
The second derivatives of the potential give the dimensionless surface mass density, $\kappa=\nabla^2\psi/2$, of the deflector, also called the convergence.

Importantly, we note that Eq.~\ref{eq:lens_eq} is, for most choices of deflection field \vec{\alpha}, non-linear with respect to the two coordinate variables of $\vec{\theta}$, even when the source position and gravitational potential are known. As a consequence, no root-finder algorithm is guaranteed to find a complete set of solutions if not initialised close to a suitable location in parameter space.

\subsection{Linear approximation \label{ssec:linear_lensing}}

Equation \ref{eq:lens_eq} allows us to determine the true (i.e. unlensed) position of a source given its lensed image and a model of the lensing mass. This inverse operation, however, requires solving an equation that may not have a general closed-form solution. Moreover, in the strong lensing regime, the equation is likely to have multiple solutions corresponding to duplicate and magnified images. For this reason, efficient modelling methods do not typically solve the lens equation for \vec{\theta}, but rather evaluate analytical profiles on a grid transformed by Eq.~\ref{eq:lens_eq} or instead describe surface brightness directly on a grid in the source plane. The latter option is made possible with the linear framework developed in \cite{WarrenDye2003}.

The lensing phenomenon smoothly defined by Eq.~\ref{eq:lens_eq} can be formulated as a linear lensing operator \lensingop acting on an image \source of source intensity values. In this formulation, \source is a flattened vector of the underlying two-dimensional image. The action of \lensingop on \source results in a lensed version of \source. We note that while the mapping between the image and source planes is formulated as an algebraic linear operation, the intrinsic non-linear nature of lensing (i.e. image multiplicity) is retained. Additional observed features, such as light from the  deflector, instrumental blurring, and other contaminants along the line of sight, can be straightforwardly included in a linear fashion. We can thus write a model \model for strong lens imaging data as
\begin{align}
\label{eq:general_model}
    \model = \convop\left(\lensingop\,\source + \lens + {\rm other\ components}\right) + {\rm noise}\ ,
\end{align}
where \lens is the surface brightness of the lens galaxy (as a flattened vector), and \convop accounts for seeing effects and/or blurring by the instrumental point spread function (PSF). Other components of the model can include, for instance, the multiple images of a background quasar or satellite galaxies near the main deflector. Several pixel-based methods to model strong lenses are based on Eq.~\ref{eq:general_model}, such as the semi-linear inversion method \citep[SLI,][]{WarrenDye2003} and the works of \cite{VegettiKoopmans2009} and \cite{TagoreKeeton2014}. Software implementations can be found in \textsc{GLEE} \citep{SuyuHalkola2010} and \textsc{AutoLens} \citep{Nightingale2018}. The present work relies on an independent implementation of the pixel-based formalism. Furthermore, our primary focus is on the source reconstruction problem at fixed lens model (i.e. fixed lensing operator \lensingop). The task of optimising \lensingop\ -- ideally using a similarly flexible method -- along with light components requires several iterations, as the lens model behaves non-linearly. We give in Sect.~\ref{ssec:mass_sampling_mock_HST}, however, an example of computing the lens model using the standard approach of sampling analytical parameters, solving the pixel-based source reconstruction at each iteration.

We note that an additional hyper-parameter\footnote{Depending on the method, other hyper-parameters can be introduced (e.g. controlling the mapping when using adaptive gridding strategies).} is implicitly assumed in Eq.~\ref{eq:general_model}, due the pixelated nature of the linear approximation: the choice of source plane resolution, or alternatively the ratio of data to source resolution. For clarity, we introduce \pixsizeratio as
\begin{align}
    \label{eq:def_rpix}
    \pixsizeratio \equiv \frac{{\rm source\ pixel\ size}}{{\rm image\ pixel\ size}}\ .
\end{align}
As noted in \cite{Suyu2006}, the source pixel size can be roughly related to the average lensing magnification. As it will be introduced later in the section, a large \pixsizeratio introduces more degrees of freedom in the model, which may not be constrained by the data alone. The choice of external constraints informed by the properties of the source light distribution allow for larger \pixsizeratio values while keeping the modelling accurate. The technique introduced in \citetalias{Joseph2019} and the further improvements we provide in this work have been designed to allow for exactly this.

\subsection{An underconstrained problem}

Given an observation \data, our goal is to solve Eq.~\ref{eq:general_model} by finding the most suitable \source and \lens\ -- ignoring any other luminous components -- for a known \convop in the presence of noise. This amounts to minimising the difference between the data and our model \model, commonly quantified by the function
\begin{align}
\label{eq:datafid}
    \likelihood{\data,\,\source,\,\lens} = \frac12 \normtwo{\data - \model}^2 \ ,
\end{align}
where $\normtwo{\,\cdot\,}$ is the Euclidean norm. Being strictly convex and differentiable, this norm is useful as a data fidelity term in optimisation algorithms.

Although mathematically simple, the linear formulation as described in Sect. \ref{ssec:linear_lensing} leads to a highly underconstrained problem. Data pixel coordinates mapped to the source plane according to Eq.~\ref{eq:lens_eq} do not coincide anymore with a cartesian coordinate grid. The choice of how to discretise the source plane can introduce errors on the reconstruction. A common technique to reduce such errors, which we adopt, is to increase the resolution of the grid on which we reconstruct the source galaxy (i.e. increase \pixsizeratio). Combined with constraints that promote smoothness (see Sect.~\ref{ssec:theory_reg}), this allows for more accurate conservation of surface brightness but also introduces more unconstrained degrees of freedom, as a subset of source pixels may not be mapped to any data pixel.

To solve the underconstrained problem, it is therefore necessary to somehow restrict the space of possible solutions. We can do this by leveraging independent knowledge regarding the unknowns that is complementary to the imaging data itself. External knowledge of this sort serves to \emph{regularise} the problem, and we denote the restrictive action of regularisation on a model component as \prior{\,\cdot\,}. The augmented minimisation problem can then be written
\begin{align}
\label{eq:general_problem}
    \argmin_{\source,\ \lens}\, \left[ \likelihood{\data,\,\source,\,\lens} + \prior{\source} + \prior{\lens} \right] .
\end{align}
Equation \ref{eq:general_problem} can also be understood in a Bayesian context, with $\mathcal{L}$ corresponding to the log-likelihood distribution and $\mathcal{P}$ corresponding to log-prior probabilities based on, for example, physical arguments or a previous solution to the problem.

\begin{figure*}[t!]
    \centering
    \includegraphics[width=\linewidth]{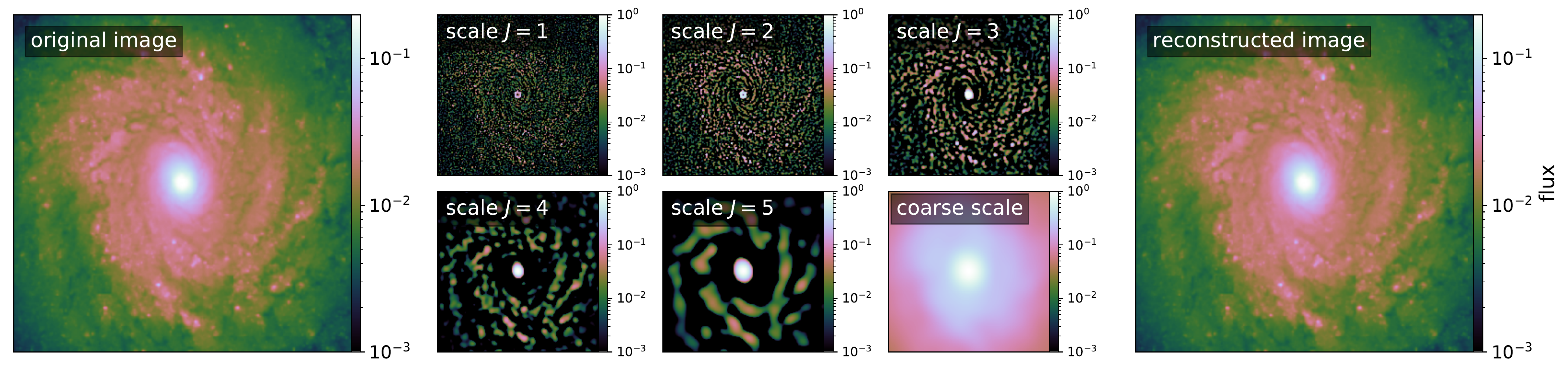}
    \caption{Multiscale decomposition using the starlet transform. \textbf{Left panel}: original image, similar to source galaxies in Sect.~\ref{sec:hst_data} and \ref{sec:elt_data}. \textbf{Middle panels}: scales of the starlet decomposition using 6 spatial scales, and normalised (${\rm max}=1$) to enhance the contrast. In such a decomposition, a feature of a given scale is ``brighter'' in the corresponding image in the decomposition. Hence, keeping only highest pixels in each starlet scale filters out any insignificant signal over a large spectrum of spatial extents. All scale images have the same dimensions as the input image. \textbf{Right panel}: reconstructed image from all the decomposition scales, through co-addition of all decomposition scales. \href{https://github.com/aymgal/SLITronomy-papers/blob/master/paper_I/visualize_starlet_transform.ipynb}{\faGithub}}
    \label{fig:starlet}
\end{figure*}

\subsection{Regularisation strategies \label{ssec:theory_reg}}

Numerous strategies to regularise source galaxy reconstruction have been explored in the literature. A fundamental regularisation shared by all pixel-based techniques arises implicitly, as noted by \cite{TagoreJackson2016}, in choosing to express surface brightness on a pixelated grid rather than continuously. However, for complex sources, such as star-forming galaxies, still further regularisation is required.

The maximum entropy regularisation method (MEM) was used in \cite{Wallington1996} to recover astronomical sources from noisy images. The goal of MEM is to maximise the total Shannon-Jaynes entropy $\prior{\source} \propto \sum_k s_k\,\ln s_k$ of the source pixels $\source = \{s_k\}$. This constraint accommodates incomplete data and ensures positivity of the solution, but it finds its minimum when all pixels $s_k$ share the same value. This contradicts the goal of accurately reconstructing sources that show a large dynamical range, which is often the case in practice. In addition, the inversion problem becomes non-linear. To retain linearity, the similar $\ell_2$ norm regularisation, i.e. $\prior{\source} \propto \sum_k s_k^2$, is often used \citep[e.g.][]{WarrenDye2003,Suyu2006}, although non-negativity is no longer guaranteed.

Higher-order choices such as gradient and curvature regularisations are designed to penalise large first and second spatial derivatives, respectively, of the reconstructed image \citep{Suyu2006}. The smoothness of the solution increases with increasing regularisation order. Methods based on (irregular) adaptive grids have used algorithm-specific forms for \prior{\source} \citep[e.g][]{VegettiKoopmans2009,NightingaleDye2015}, and still other families of regularisation have been applied in a Bayesian framework \citep{BrewerLewis2006}.

Machine learning techniques, such as neural networks, are also effective regularisation methods. More specifically, given a suitable dataset of galaxies that are close enough to real observations, supervised learning techniques are expected to extract complex structural information about the light distribution, which subsequently regularises the model \citep[see e.g.][]{Morningstar2019,Pearson2019,Chianese2020}.

Each method has its advantages and disadvantages, and one may be more suitable than another depending on the quality of the data, the intrinsic morphology of the source and the science case. As simple convex functions, MEM and $\ell_2$ norm regularisations have analytic derivatives that are useful in gradient descent algorithms but may overfit the data. Higher-order expressions reduce the risk of overfitting but may be time-consuming to compute and fail to properly account for local variations in the source. Moreover, regularisation from deep learning can be severely limited by the realism of training datasets. We describe our choice of regularisation in detail in Sect. \ref{ssec:sparse_reg}.

\section{The Sparse Linear Inversion Technique (SLIT) \label{sec:slit}}

This work aims to improve and expand the source inversion framework introduced in \citetalias{Joseph2019}, which pioneered the use of wavelets and sparse regularisation in strong lens modelling. Here we review and update the original problem and solution as implemented by, putting the regularisation method more into context.

\subsection{Data model \label{ssec:slit_model}}

The original Sparse Linear Inversion Technique (SLIT) algorithm introduced the following linear problem to solve \citepalias{Joseph2019}:
\begin{align}
\label{eq:slit_model}
    \model = \convop\,\lensingop\,\source + \lensconv + \noise,
\end{align}
where \model, \convop, \lensingop\footnotemark, and \source retain their meanings from Eq.~\ref{eq:general_model}, ignoring other components. Here $\lensconv \equiv \convop\lens$ is the galaxy lens light convolved with the instrumental PSF. Solving for \lensconv instead of the deconvolved \lens removes a potentially costly step in the modelling process and is anyway not needed for many scientific applications. We take \noise to be white Gaussian noise, although it is possible to treat additive Poisson noise as well. For a fixed lens mass model, that is, known \lensingop, the task is to jointly solve for the deconvolved and denoised source surface brightness in the source plane \source and the convolved lens surface brightness in the image plane \lensconv. The algorithm can also optimise for \source alone when the lens light has been subtracted beforehand.

\footnotetext{We have dropped ``$\kappa$'' from the original notation $\mat{F_\kappa}$ in \citetalias{Joseph2019}, since we do not build the lensing operator from the convergence map $\kappa$, but rather via deflection angles through ray-tracing.}

The original paper distinguished two versions of the algorithm: \texttt{SLIT} when solving only for \source and \texttt{SLIT\_MCA} when solving for both \source and \lensconv. The latter is based on the framework of morphological component analysis (MCA) developed for blind source separation \citep{Starck2005,Bobin2008, Joseph2016}. The technique seeks to disentangle superimposed signals based on their morphological properties when represented in a certain basis. This is relevant in strong lensing, as the deformed source galaxy light is often blended to some extent with the lens galaxy light. In \texttt{SLIT\_MCA}, it is assumed that \source and \lensconv can each be more sparsely represented in their respective planes by a particular wavelet transform, allowing the two components to be separated.

\begin{figure*}
    \centering
    \includegraphics[width=\linewidth]{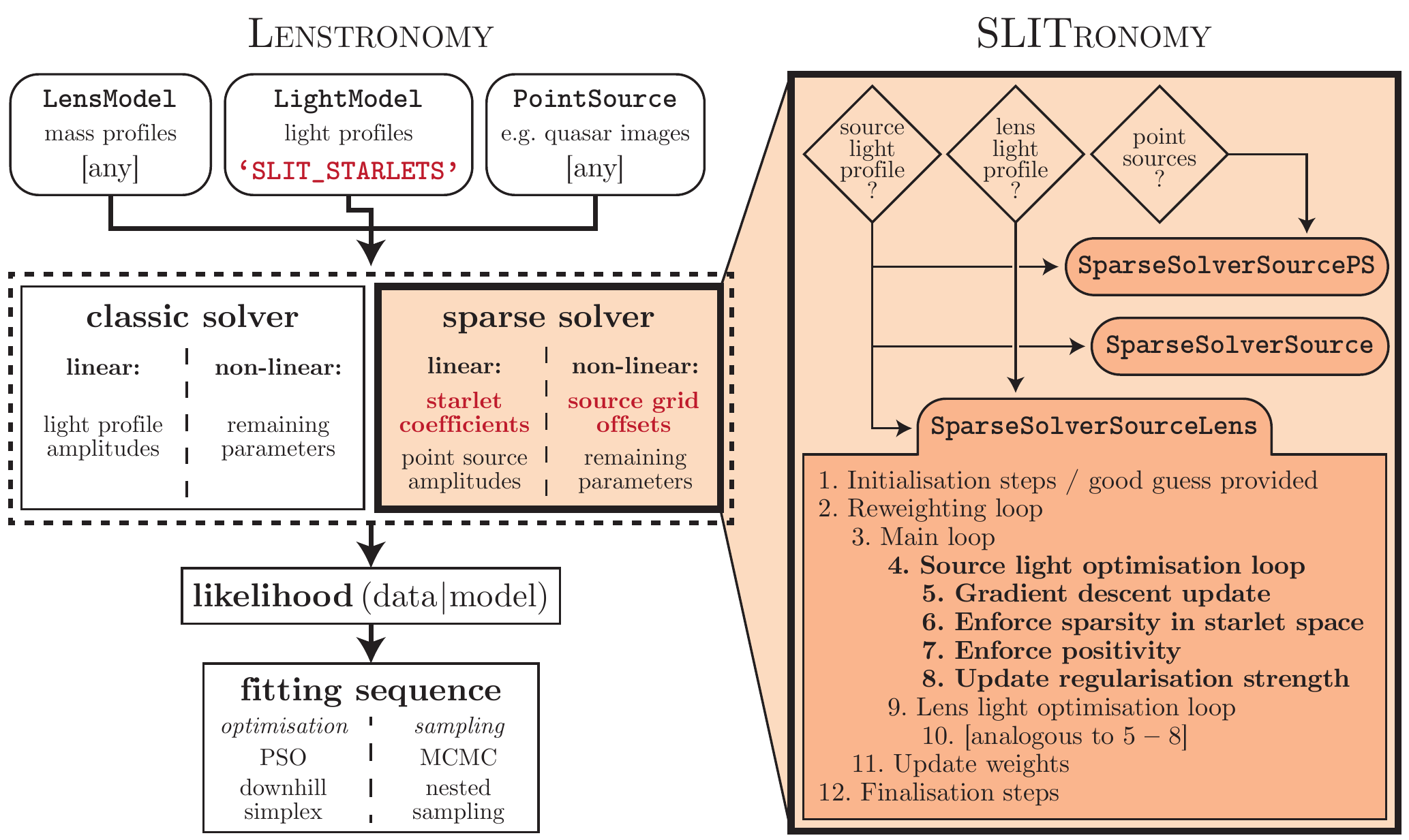}
    \caption{Integration of \slitro into \lenstro. The global architecture of \lenstro is left unchanged. When the \texttt{'SLIT\_STARLETS'} profile is used (for modelling source or lens surface brightness), the image likelihood is computed through the sparse solver of \slitro, instead of the usual solver. Changes in \lenstro are indicated in red: a new light profile \texttt{'SLIT\_STARLETS'} has been introduced; linear amplitudes (originally defined for analytical profiles) are replaced by a sparse optimisation of individual coefficients in starlet space; source grid coordinate offsets are added as non-linear parameters, jointly optimised with e.g. lens model parameters. The integration of point source modelling is also illustrated. See Sect. \ref{sec:slitronomy} for details.} \label{fig:lenstro_slitro_integration}
\end{figure*}

In this paper, we do not explicitly distinguish between the two algorithms. This is to emphasise that the new implementation is a single tool that adapts to the intended use. Where appropriate, we simply specify whether we solve for the source galaxy alone, for both the source and lens galaxies, or for the source galaxy and point sources (see Fig.~\ref{fig:lenstro_slitro_integration} and Sect. \ref{sssec:ps_support} for details of point source modelling).

\subsection{Sparsity and starlet regularisation \label{ssec:sparse_reg}}

As summarised in Sect.~\ref{ssec:theory_reg}, recent pixel-based reconstruction techniques make use of some variation of an adaptive grid for the source plane. In our work, we choose not to use an adaptive gridding scheme but instead a uniform source grid with resolution much higher than that of the data. This is for two main reasons: (1) our regularisation method is intrinsically multi-scale as we work in the wavelet space; (2) pixelated lensing operations and wavelet transforms are faster when applied on a uniform grid. We further allow the source plane grid to have the minimal size it can have given the data size, lens model, and (optionally) masking strategy.

Most regularisation strategies share the common difficulty of deciding how to set the strength of regularisation. The strength is controlled by a Lagrange parameter $\lambda$ within $\mathcal{P}$ that balances its importance relative to the data fidelity term $\mathcal{L}$ (cf. Eqs.~\ref{eq:general_problem} and \ref{eq:prior_S+G}). A further complication is that $\lambda$ changes the effective number of degrees of freedom in a way that is difficult to quantify. This issue has motivated numerous uses of Bayesian arguments to objectively set $\lambda$ \citep{Suyu2006,BrewerLewis2006,Dye2008,VegettiKoopmans2009}. Except for simple forms of regularisation, however, such as quadratic expressions, computing the Bayesian evidence often incurs significant computational overhead, as it requires integrals over a large parameter space volume.

We rely instead on sparsity in the wavelet domain to meaningfully set $\lambda$ based on noise properties of the data. Sparse image reconstruction is a well-studied framework that has been used in a wide variety of astrophysical applications, such as the processing of weak-lensing \citep[e.g.][]{Lanusse2016,Peel2017} and radio interferometric data \citep[e.g][]{Pratley2018}, blind source separation of optical and radio sources \citep[e.g.][]{Joseph2016,Jiang2017}, and recently in combination with deep learning \citep[e.g][]{Sureau2019}. 

The success of wavelets in such applications stems from their multi-resolution property via the wavelet transform. Similarly to the way a signal is decomposed into its component frequencies by a Fourier transform, a wavelet transform decomposes a signal into both its frequency and spatial components. Figure \ref{fig:starlet} illustrates the multi-resolution decomposition of an image of a galaxy. A wavelet transform is defined by its basic functional wavelet form, which must obey certain mathematical properties. The particular choice of wavelet depends on the problem at hand. In this work we use the starlet transform \citep{Starck2007}, which is isotropic and undecimated, and has been shown to be suitable for treating astronomical images.

The starlet transform of an image returns $J$ new images of equal size, each corresponding to a convolution of the original by a filtering kernel of a different scale. The $j$th kernel $j\in\{1,...,J-1\}$ amplifies features of the image on scales of $2^j$ pixels (see decomposition scales panels in Fig.~\ref{fig:starlet}) while preserving spatial locations. The final image in the decomposition corresponds to coarse-grained smoothing similar to a Gaussian filtering (see coarse scale panel in Fig.~\ref{fig:starlet}). One sets the number of scales in the transform by hand, the maximum value being limited by the number of pixels $n_{\rm pix}$ on the side of the image: $J_{\rm max}=\lfloor\log_2{n_{\rm pix}}\rfloor$. We note that the number of decomposition scales might also be optimised further for a specific system. We may explore this possibility in future work. Following standard notation \citepalias[and that of][]{Joseph2019}, we write the linear starlet transform operator as\footnotemark\ \starletsop and its inverse as \starletsinvop. The starlet transform of an image $\mat{x}$ yields coefficients $\alpha_\mat{x}=\starletsop\mat{x}$, and $\mat{x}$ is recovered with $\mat{x}=\starletsinvop \alpha_\mat{x}$.

\footnotetext{Formally one should write \starletsopexplicit{\{\source,\lensconv\}} and \starletsinvopexplicit{\{\source,\lensconv\}}, where the subscript in \starletsopexplicit{\{\source,\lensconv\}} indicates if the operator is meant to be applied on the source or lens galaxy, with the corresponding number of scales \Jscales{\{\source,\lensconv\}}. To avoid cluttered expressions, we drop this subscript, and depending on which component the operator is applied on (\source or \lensconv), the operator is the starlet operator whose number of decomposition is scales \Jscales{\source} or \Jscales{\lensconv}, respectively. That is, $\starletsop\source\equiv\starletsopexplicit{\source}\source$ and $\starletsop\lensconv\equiv\starletsopexplicit{\lensconv}\lensconv$.}

We regularise the reconstruction of both lens and source galaxies by applying a sparsity constraint on their starlet coefficients. This is motivated by the observation that astronomical images, including the light and mass distributions of galaxies, can be expressed with high fidelity using only a small number of coefficients in starlet space \citep[e.g][]{Lanusse2016,Joseph2016,Peel2017}. This property therefore defines a suitable regularisation for the underconstrained problem described by Eq.
~\ref{eq:general_problem} in a model-independent way. The sparsity of a signal is enforced by minimising its $\ell_0$ or $\ell_1$ norm, in our case in the starlet domain. The former is generally preferred but difficult to use in practice, because it is not convex. \slit uses the $\ell_1$ norm, as it still promotes sparsity and being convex, guarantees convergence. We can now write the final expression of our priors as
\begin{align}
\label{eq:prior_S+G}
     \prior{\mat{x}} = \lambda\,\normone{\Wweights{\mat{x}}\odot\starletsop \mat{x}} + \imath_{\geq0}\left(\mat{x}\right),
\end{align}
where \mat{x} stands for either \source or $\lens_\convop$. We have introduced weights \Wweights{\mat{x}} that adjust the starlet coefficients via an element-wise product $\odot$. This is necessary because the $\ell_1$ norm (unlike the $\ell_0$ norm) is known to bias the amplitude of the reconstructed signal, but this can be mitigated by an appropriate reweighting \citep{Candes2007}. We give more details in Appendix \ref{app:reweighting} on the implementation of the $\ell_1$-reweighting step. To avoid unphysical solutions, such as those containing negative pixels, we include as well the positivity constraint $\imath_{\geq0}(\,\cdot\,)$, which goes to $+\infty$ if any coefficient of its argument is negative, set to zero otherwise, and is self-consistently folded in during optimisation.

Solving the convex minimisation problem of Eq.~\ref{eq:general_problem} subject to the constraints given by Eq.~\ref{eq:prior_S+G} is a difficult task because the $\ell_0$ and $\ell_1$ norm are not differentiable everywhere. A large family of methods to solve such minimisation problems rely on splitting the different constraints and applying them through more tractable mathematical operators. These are called proximal operators associated with each constraint and which can offer convergence guarantees depending on the algorithm and number of constraints. The proximal operator of the $\ell_1$ norm is the soft-thresholding operation. The proximal operator of the $\imath_{\geq0}(\,\cdot\,)$ corresponds to setting to zero all negative values. These operators are often applied after the step corresponding to a gradient descent update, which is the case in our algorithm.

As described above, the starlet transform gives a multi-resolution decomposition of the source being modelled. It encodes the spatial locations of features across many scales simultaneously. This is another advantage compared to classical regularisation strategies, like those introduced in Sect.~\ref{ssec:theory_reg}. On one hand, classical strategies are ``global'' in that every image pixel contributes equally to the regularisation (through summation over the pixels), whereas starlet regularisation retains spatial locations. On the other hand, while MEM and the $\ell_2$ norm are single-pixel measures, and gradient and curvature correlate flux over a few pixels, starlet regularisation incorporates correlations over a much wider range of scales, resulting in a higher fidelity reconstruction (see Fig.~\ref{fig:starlet}).

\subsection{Regularisation strength}

The value of the Lagrange parameter $\lambda$ generically regulates the importance of regularisation compared to accuracy in reproducing the observation. A higher $\lambda$ favors a smoother reconstruction at the cost of a worse fit to the data, whereas a lower $\lambda$ gives more flexibility in matching the data at the pixel level but increases overfitting. In the context of sparse regularisation through the $\ell_0$ or $\ell_1$ norm, $\lambda$ also has a clear interpretation: it effectively sets the statistical significance of the reconstruction in units of the noise, provided that the noise covariance is known and properly propagated to the starlet domain. We denote the propagated noise as $\sigma_{\starletsinvop}$ \citepalias[see][for its computation]{Joseph2019}, hence $\lambda\equiv\lambda'\sigma_{\starletsinvop}$ for a positive scalar $\lambda'$. For instance, $\lambda'=3$ leads to a reconstruction at ``$3\sigma$~significance''. This understanding limits the range of interest for $\lambda'$ values to typically $\ge3$. In the algorithm, we start with a large $\lambda'$ -- automatically estimated from $\sigma_{\starletsinvop}$ -- and reduce it at each iteration until it reaches the target value. It is then held fixed for a small number of remaining iterations (typically 5).

We set $\lambda'=3$ throughout this paper, as we have found it generally gives the smallest image plane residuals in our experiments, with no sign of overfitting. We also allow for a larger threshold applied specifically to the smallest starlet decomposition scale, to prevent spurious isolated pixels from entering the solution. As detailed in Sect.~\ref{ssec:sparse_reg}, the first starlet scale contains features with a spatial extent of $(2^j)_{j=0}=1$ source pixel, which are likely to be noise. As a consequence, we set $\lambda'=5$ for the smallest starlet scale only. We note that alternative solutions may exist, such as replacing the first starlet scale by a more suitable wavelet transform chosen such that isolated pixels are more penalised by the sparsity constraint \citep[e.g.][using a hybrid Battle-Lemari\'e+starlet transform]{Lanusse2016}.

A careful handling of the noise in the context of sparse modelling thus gives a simple way to choose the regularisation strength according to the desired reconstruction significance. A systematic comparison of the Bayesian evidence \citep{Suyu2006,Dye2008,VegettiKoopmans2009} is not necessary in our framework, which saves considerable computation time when solving the full \slit problem (Eq.~\ref{eq:slit_problem}). We refer to \citetalias{Joseph2019} for further details concerning the computation of noise levels and their propagation through \slit operations.

\section{From \slit to \slitro \label{sec:slitronomy}}

\begin{figure*}
    \centering
    \includegraphics[width=0.8\linewidth]{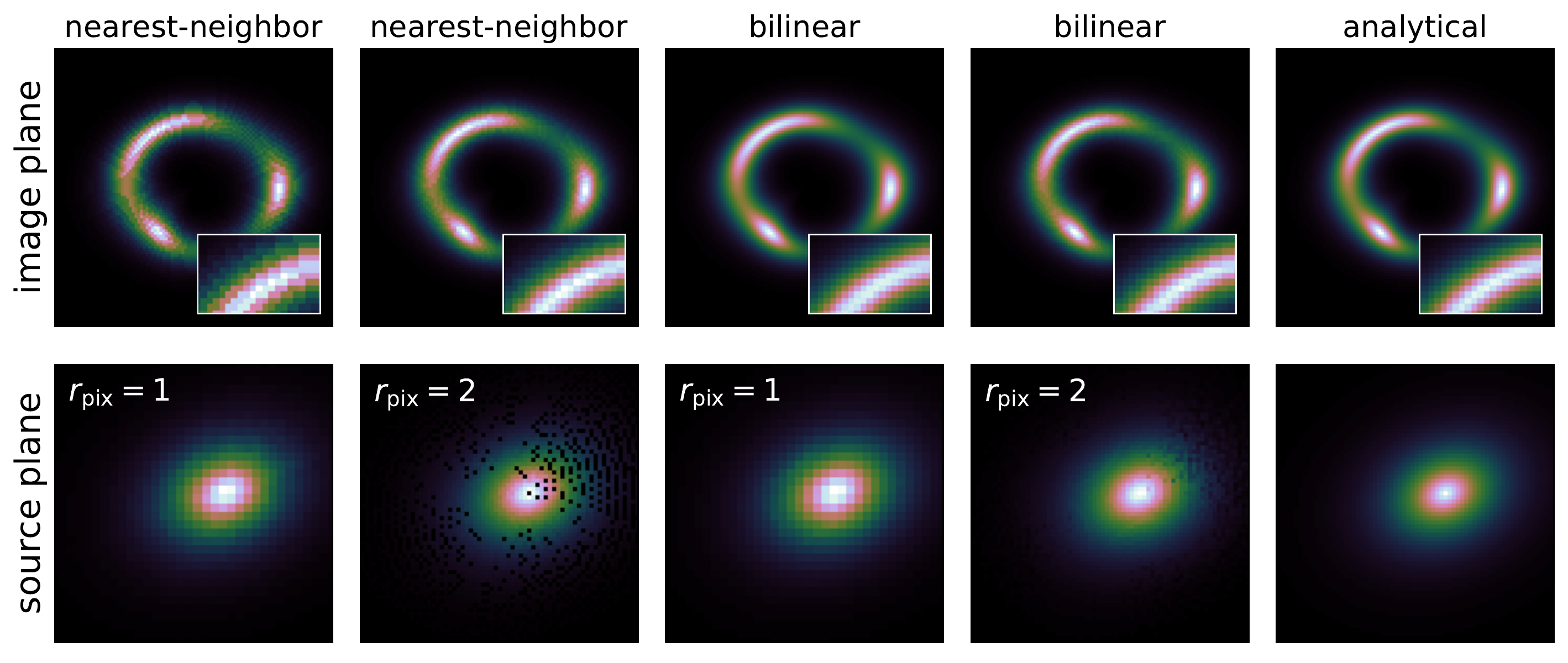}
    \caption{Illustration of a pixelated mapping between source and image planes, as implemented in \slitro. \textbf{Top row, left to right}: nearest-neighbor interpolation of source surface brightness for two grid resolutions (same as image plane or doubled), bilinear interpolation of surface brightness for the same resolutions, and parametric ``groundtruth'' using \lenstro. \textbf{Bottom row, left to right}: source plane corresponding to the top row panels, for different data pixel size to source pixel ratio \pixsizeratio. Dark isolated pixels in source plane are not mapped to any image plane pixel, hence not constrained by imaging data. Our reconstruction technique is able to fill these ``missing pixels'', through sparse regularisation and multiresolution property of wavelets. \href{https://github.com/aymgal/SLITronomy-papers/blob/master/paper_I/visualize_lensing_operator.ipynb}{\faGithub}}
    \label{fig:lens_mapping}
\end{figure*}

The original \slit algorithm presented in \citetalias{Joseph2019} is available as an open-source python code\footnote{\url{https://github.com/herjy/SLIT}}, packaged with example scripts for illustrating its different features. There are two main difficulties with the code that we have sought to improve in this work. The first is that the computation time required for source reconstructions, and consequently joint source-lens reconstructions, is too large for tractable applications to real data. The second is that it is not practical to initialise the algorithm with results from the literature (e.g. lens model parameters), due to differences in coordinate systems and conventions. This is true for using the algorithm as a standalone tool, as well as in a workflow where model parameters first get optimised through analytical methods, then further refined using the \slit algorithm.

We address these issues by introducing \slitro\footnote{\url{https://github.com/aymgal/SLITronomy}}, our revamped implementation of the \slit algorithm that is significantly more efficient than its predecessor. It can be used as a plug-in to the open-source gravitational lens modelling software \lenstro\footnote{\url{https://github.com/sibirrer/lenstronomy}} \citep{Birrer2018lenstro}. The framework of \lenstro allows us to access a number of practical and tested features that we exploit in our new development. The support of our code within \lenstro allows our sparse lens inversion method to be easily used by people already familiar with this software.

In the rest of this section, we describe a subset of key improvements and new features that we have brought to the original \slit package. We note that not all features described in the following subsections are required for the results presented in Sects. \ref{sec:hst_data} and \ref{sec:elt_data}.

\subsection{Analysis and synthesis formulations}

One can rewrite the general minimisation problem of Eq.~\ref{eq:general_problem} with the \slit model from Eq.~\ref{eq:slit_model}, and the regularisation strategy defined by Eq.~\ref{eq:prior_S+G}. At this stage, there are two ways to solve the problem, depending on which variables one chooses to optimise. The \emph{synthesis} formulation solves the problem in the transformed domain, meaning that the actual variables are starlet coefficients of the source $\alpha_\source$ and lens galaxies $\alpha_\lens$. The output source and lens images, \source and \lensconv, are then obtained by applying a final inverse starlet transform. In this case, one tries to find the sparsest solution as measured in starlet space. On the other hand, the \emph{analysis} formulation of the problem solves for \source and \lensconv in direct space. In this case, one seeks a solution in the data space that simultaneously has a sparse representation in starlet space\footnote{We note that the analysis and synthesis formulations are strictly equivalent only if $\starletsop\starletsinvop$ is the identity. As the starlet functions form an overcomplete basis, this condition is not met for the starlet transform.}.

In the original \slit implementation, the model of Eq.~\ref{eq:slit_model} is optimised using the synthesis formulation. The main advantage of the synthesis formulation relies in the fact that the proximal operator of $\normone{\alpha_\mat{x}}$, where $\alpha_\mat{x}$ are starlet coefficients of an image \mat{x}, is exactly the soft-thresholding operator. This enables the use of simple and efficient algorithm such as the forward-backward algorithm \citep{Combettes2005}, without any approximation. In our implementation, we chose to solve the problem in the analysis formulation. The problem of Eq.~\ref{eq:general_problem} is then written explicitly as:
\begin{align}
\label{eq:slit_problem}
    \nonumber
    \argmin_{\source,\ \lensconv}&\, \frac12 \normtwo{\data - \convop\,\lensingop\,\source - \lensconv}^2 + \imath_{\geq0}\left(\source\right) + \imath_{\geq0}\left(\lensconv\right) \\
    &+ \lambda\,\normone{\Wweights{\source}\odot\starletsop \source}
    + \lambda\,\normone{\Wweights{\lens}\odot\starletsop \lensconv} \ ,
\end{align}
where $\odot$ stands for the element-wise product, and we recall $\lambda\equiv\lambda'\sigma_\starletsinvop$.

The difference with the synthesis formulation is that the variables are now \source and \lensconv, that is, galaxy images in direct space. First, the number of effective free parameters (before regularisation is applied) is largely reduced when solving in direct space: it is simply the number of pixels $N^2$ as opposed to $J\times N^2$ for $J$ decomposition scales. In this regard, optimisation in the analysis formulation is simpler and more stable. Second, the positivity constraint is simple to apply in direct space, because it does not require sub-iterations like in transformed space. Third, some specific features, such as a those similar to a dirac function, are better represented in direct space, whereas they may require many more coefficients in transformed space. One drawback of the analysis formulation is that the sparsity constraint is not mathematically described by the exact soft-thresholding operation, compared to the synthesis formulation. In practice, however, the process of sequentially applying the starlet transform, the soft-thresholding operation, and then the inverse starlet transform does not prevent the algorithm from converging. There exist algorithms capable of solving Eq.~\ref{eq:slit_problem} without this approximation \citep[e.g][]{Vu2011}, but their implementation is left for a future version.

As an additional improvement to the overall computation time, we use an optimised implementation of the starlet transform through the python package \texttt{pySAP}\footnote{\url{https://github.com/CEA-COSMIC/pysap}} \citep{Farrens2019}.

\subsection{Lensing operator \label{ssec:slitro_lensing_op}}

In \slitro, the implementation of the lensing operator (\lensingop in Eq.~\ref{eq:slit_model}) has been improved for efficiency. When represented as a matrix, this operator is known to be extremely sparse, typically containing fewer than 1\% of non-zero entries. In the original implementation, only non-zero entries were saved in memory to carry out lensing operations. In our new implementation, we use a faster and more memory efficient technique to build the operator based on its matrix representation and provided by the \texttt{scipy.sparse}\footnote{\url{https://www.scipy.org}} python package. In addition to storage optimisation, the matrix representation allows us to carry out the $\lensingop\,\source$ operation and its inverse as plain matrix-vector products, which is considerably faster. We give quantitative measures of improved computation times in Table \ref{tab:main_speedups} corresponding to a configuration of $100\times100$ data pixels with twice the resolution in the source plane on a recent personal computer. This setup is typical, for example, of imaging data in the WCF3/F160W band of HST. To give an idea of the total computation time needed to solve problem \ref{eq:slit_problem}, we also compare solving for \source and $\source+\lensconv$ for the same configuration.

These speed improvements become crucial as soon as one wants to solve the more general problem of optimising the lens model in addition to the source reconstruction for fixed lens model (see Sect.~\ref{ssec:mass_sampling_mock_HST}). It is also highly beneficial for the more difficult problem of deblending and modelling both the source and lens galaxies. More detailed benchmarks are shown in Appendix~\ref{app:benchmarks} for various settings of the algorithm and data sizes.

\begin{table*}[htbp!]
\caption{Speed up improvements of the new \slitro implementation, compared to the original SLIT code. These numbers are valid for a the typical configuration of $100\times100$ data pixels, with $\pixsizeratio=2$ for source resolution, and $\Jscales{\source}=6$ starlet scales. It can be for instance the modelling of a strong lens system imaged by HST in F160W band. \label{tab:main_speedups}}
\renewcommand{\arraystretch}{1.3}
\centering
\begin{tabular}{l c c c}
\hline
operation & SLIT & \slitro & speedup \\
\hline\hline
computation of the lensing operator, \mat{F} & 1.7 s & 6 ms & 300 \\
lensing of an image, $\lensingop\mat{x}$ & 200 ms & 450 $\mu$s & 4'500 \\
delensing of an image, $\lensingop^\top\mat{y}$ & 400 ms & 120 $\mu$s & 3'500 \\
solve for source light, $\source$ & $\sim16$ min & $\sim 1$ s & 1'000 \\
solve for source and lens light, $\source+\lensconv$ & $\sim 2$ h & $\sim 30$ s & 240 \\
\hline
\end{tabular}
\end{table*}

\subsection{Pixelated source surface brightness}

The lens equation (Eq.~\ref{eq:lens_eq}) gives the mapping between image plane coordinates and source plane coordinates, provided the deflection angles are known. When considering the surface brightness of galaxies as an ensemble of pixels, the lens equation is discretised, and one defines grids of pixels for the image plane and source planes. These grids need not necessarily be rectangular, but we only consider regular Cartesian grids in this work. When ray-tracing image plane pixels to the source plane through Eq.~\ref{eq:lens_eq}, the resulting coordinate is not necessarily centered on a source plane pixel. A simple way to deal with this while keeping a Cartesian coordinates for both planes is to interpolate over neighboring source pixels to compute the values of the ray-traced image pixels. In our formalism, we consider this interpolation step as included in the lensing operator \lensingop.

In the original \slit, a nearest-neighbor interpolation scheme was chosen for its simplicity of implementation, relying entirely on the regularisation (Eq.~\ref{eq:slit_problem}) to fill unconstrained pixels. With \lenstro's new implementation of the lensing operator as described above, more complex interpolation schemes now do not significantly increase the time required to build the matrix \lensingop. We have thus implemented a bilinear interpolation over source plane pixels similar to the one described in \cite{TreuKoopmans2004}. In order to demonstrate visually the advantage of bilinear interpolation, we show in Fig.~\ref{fig:lens_mapping} the difference between the two types of interpolation for two different source plane resolutions. The bilinear interpolation leads to a much smoother light distribution that is closer to the true profile computed from analytical ray-tracing.

Another interesting observation is worth noting in Fig.~\ref{fig:lens_mapping}. When the source plane resolution is significantly higher than the data resolution (twice, in this example), one can see ``holes'' in the source light after applying the de-lensing operator. This is typical of the linear approach to strong lensing on regular grids: not all source plane pixels get mapped to by an image plane pixel. This further illustrates the need for regularisation to solve the linear problem of Eq.~\ref{eq:general_problem}, since it is intrinsically underconstrained. A regularisation based on wavelets is expected to be very efficient at filling these holes in the source plane, because wavelets provide a multi-resolution representation of the light distribution.

\subsection{Lens model optimisation}

The fact that \slitro is embedded in the \lenstro framework makes it convenient to use optimisation and sampling tools for lens model optimisation, in addition to the pixel-based reconstructions. For example, it is now simple to run a non-linear optimisation or MCMC sampling over lens model parameters with a call to the \slitro solver for each proposed sample. As mentioned in Sect. \ref{ssec:slitro_lensing_op}, the runtime of a \slitro reconstruction with a fixed mass model for a typical HST cutout image is $\sim$1 s when solving only for the source, and $\sim$30 s when solving for the source and lens light together. While solving for the source at each iteration of an MCMC process is still tractable -- we have successfully applied such a workflow -- solving for both components would be difficult given current computation speeds. Consequently, we see several possibilities to deal with this difficulty:
\begin{itemize}
    \customitem implement further speed improvements of the $\source+\lensconv$ solver. Deep learning techniques would likely accelerate the algorithm by replacing steps considered as bottlenecks \citep[see e.g.][]{Meinhardt2017,Sureau2019};
    \customitem use other non-linear solutions to explore the lens model parameter space. An interesting direction to explore is free-likelihood techniques with extreme data compression \citep[see e.g.][]{Alsing2019};
    \customitem pre-optimise the lens model using faster modeling techniques (e.g. shapelets), so only a few ``refinement'' steps are needed using starlet regularisation.
\end{itemize}
An extensive study of lens model parameter optimisation is dedicated to a future paper. We see the present work as one of the key steps needed to develop a full model for both the light and mass components.

\subsection{Offset to source grid alignment \label{ssec:source_offset}}

A difficulty shared by any pixel-based source reconstruction method is related to the definition of the source-plane coordinate grid. In particular, the alignment between image plane coordinates and source plane coordinates can be arbitrary and potentially lead to biases in lens model parameters depending on the choices made to address the problem. This is sometimes referred to as the ``discretisation bias'', and several methods have been suggested to mitigate its effect, such as randomising the source plane initialisation \citep{NightingaleDye2015} or as an additional error term in image plane\footnote{We also performed source reconstructions with the ``regridding error'' term of \cite{Suyu2009}, but found it leads to prominent features in the residuals due to overfitting.}. Mitigation strategies typically depend on the algorithm-specific implementation and its assumptions. Analytical methods are, in essence, not subject to the discretisation bias.

Our method does not involve an adaptive gridding strategy for source plane coordinates. Hence the remaining degrees of freedom left for altering its alignment with image plane coordinates are constant offsets along both main axes. We introduce two additional non-linear parameters, $\delta_{\source, x}$ and $\delta_{\source, y}$, that characterise deviations from perfect alignment between the image-plane and source-plane coordinate grids. These two parameters are included in the set of non-linear parameters defined by other model components. Best-fit values and posterior distributions are therefore consistently estimated during optimisation or sampling.

\subsection{Point source modelling \label{sssec:ps_support}}

In the context of lensed point sources, the multiple images of background quasars are often seen as just the imprint of the PSF superimposed on the light of their host galaxy. There are two ways to model these features. One way is to increase the image plane resolution to the point that it is possible to resolve each quasar image, but this leads to extremely large numbers of source-plane pixels and quickly becomes intractable. Another simpler, faster and more effective method is to shift and scale -- i.e. (de)magnify -- the pixelated PSF at the position of each image. This is the solution implemented in \lenstro
~\footnote{Given an accurate model for the source surface brightness, the pixelated PSF can further be refined iteratively.}, and other methods use the same procedure as well \citep{SuyuHalkola2010,Auger2011}. Our model of Eq.~\ref{eq:slit_model} can be straightforwardly extended to support point-source modelling:
\begin{align}
    \label{eq:slit_model_ps}
    \mat{\model_{\rm ps}} = \convop\,\lensingop\,\source + \lensconv + \sum_i A_i\psfkernel + \noise\ ,
\end{align}
where $A_i\psfkernel$ represents the interpolated and amplitude-scaled PSF kernel at the location of the $i$-th image. An iterative approach can be employed to solve for the three unknown components \source, \lensconv and $\{K_i\}$, removing from the data two out of the three components to estimate the third one in sub-iterations, until convergence.

\section{Current data with HST \label{sec:hst_data}}

In this section, we focus on imaging data corresponding to typical images obtained with the \textit{Hubble Space Telescope} (HST). We first simulate realistic HST data to perform a detailed comparison between our pixel-based reconstruction technique and state-of-the-art analytical reconstructions. We also propose an automated process for refining the source model using both methods. We then reconstruct the source galaxy in a subset of Sloan Lens Advanced Camera for Surveys (SLACS) strong lens samples.

\subsection{Source reconstruction on simulated data \label{ssec:hst_sims}}

\subsubsection{Simulation setup \label{sssec:mock_hst}}

\begin{figure}
    \centering
    \includegraphics[width=\linewidth]{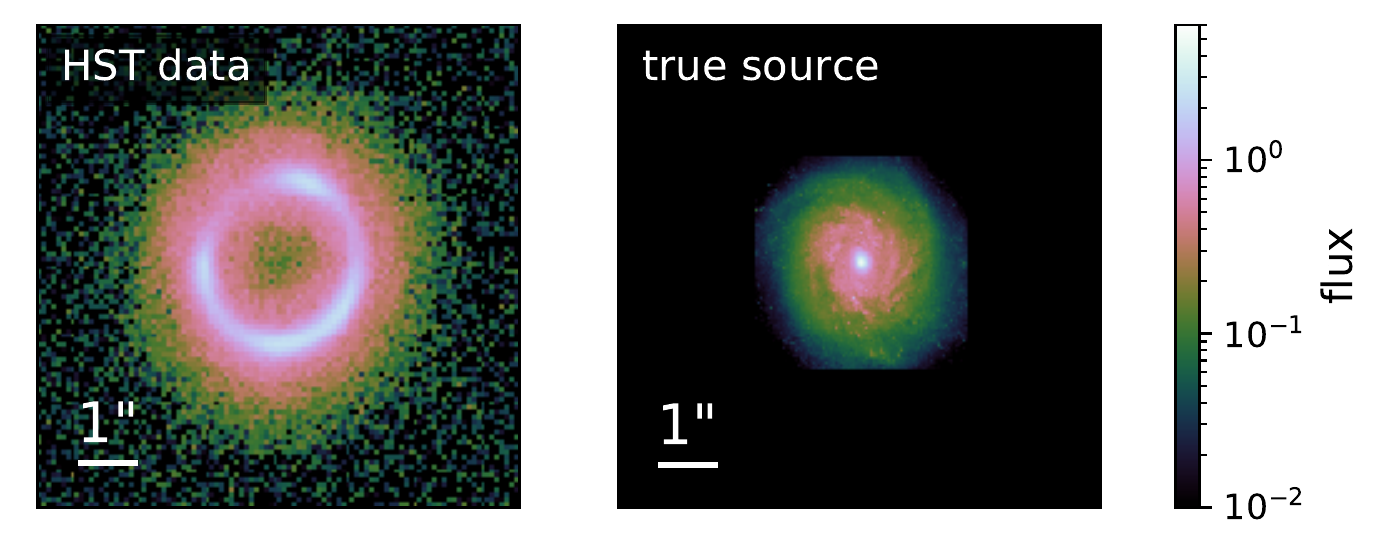}
    \includegraphics[width=\linewidth]{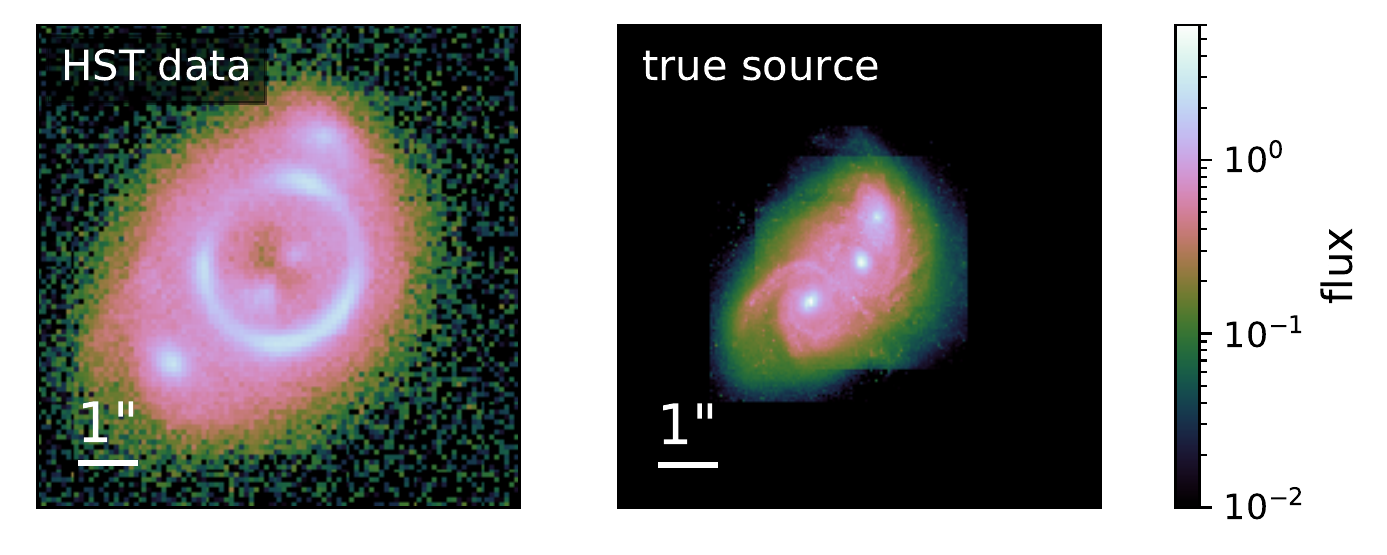}
    \caption{Simulated HST data. The lens light is assumed to be perfectly subtracted. The resolution and noise covariance mimic those of real observations with the WCF3/F160W instrument. \textbf{First row}: simulation using a single galaxy. \textbf{Second row}: more complex configuration with a composite of three close-by galaxies in the source plane. Source objects are projected on a finer grid, to match the resolution of our reconstruction technique. \href{https://github.com/aymgal/SLITronomy-papers/blob/master/paper_I/mock_source_reconstruction.ipynb}{\faGithub}}
    \label{fig:mock_HST}
\end{figure}

We use as sources the preprocessed high-resolution galaxies that were employed to generate the simulations of Time Delay Lens Modelling Challenge \citep[TDLMC, see][]{Ding2018_tdlmc,Ding2020_tdlmc}. These sources consist of high-resolution cropped images of nearby spiral galaxies from the Hubble Legacy Archive\footnote{\url{https://hla.stsci.edu/hlaview.html}}, on which isolated stars and objects in the field have been removed and the background has been subtracted. Doing so ensures that a negligible amount of noise is propagated to the image plane, the presence of which can potentially lead to biases in the source reconstruction and recovered lens model parameters. We create two types of source: either a single galaxy or two close-by galaxies simulating a merging pair, that we place at redshift $z_{\rm s}=1.2$. Source galaxy images are shown in the right column of Fig.~\ref{fig:mock_HST}.

Once the source galaxy has been selected, we use \lenstro and a built-in light profile that performs spline interpolation of the (high-resolution) source image. We then simulate the lensing effect through ray-tracing on the interpolated image, based on an analytical model for the lens galaxy mass. The main deflector is described by a singular isothermal ellipsoid (SIE) placed at redshift $z_{\rm d}=0.3$ with a central velocity dispersion of $\sigma_{\rm v}=260$~km/s and axis ratio $q_{\rm m}=0.8$. In addition, it is embedded in an external shear with moderate strength $\gamma_{\rm ext}=0.03$, misaligned by $\pi/4$ rad upwards, counter-clockwise, with respect to the lens galaxy, to simulate the influence of nearby perturbers. After ray-tracing to compute extended lens arcs, we convolve the image with the HST PSF kernel for the near-infrared F160W band, simulated with the TinyTim software \citep{Krist2011}. We choose the F160W band, because it usually gives a higher signal-to-noise for the lensed arcs, as well as better contrast between source and lens galaxies (although we ignore the lens light in this case). Typical measurement noise is then added in the form of a background Gaussian component and a Poisson component based on instrumental properties and exposure time, combined per pixel~$i$~as
\begin{align}
    \label{eq:def_noise}
    \sigma_i^2 = \sigma_{\rm bkg}^2 + \sigma_{\rm Poisson}^2 \ .
\end{align}
We assume uncorrelated noise in the above equation, as this approximation is common to CCD data.In general, correlated noise can be incorporated as well, as long as the correlations are known or properly estimated (the propagation of noise to wavelet space is left unchanged). Other instrumental settings (e.g. zero-point magnitude, read noise, etc.) are based on real HST observations. Final simulated images are shown in Fig.~\ref{fig:mock_HST}. We summarise in Table~\ref{app:tab:sim_settings} the instrumental settings and in Table~\ref{app:tab:sim_params} the values of the lens model parameters used in our simulations.

\subsubsection{Starlet source reconstruction \label{sssec:hst_starlets}}

\begin{figure*}
    \centering
    \includegraphics[width=\linewidth]{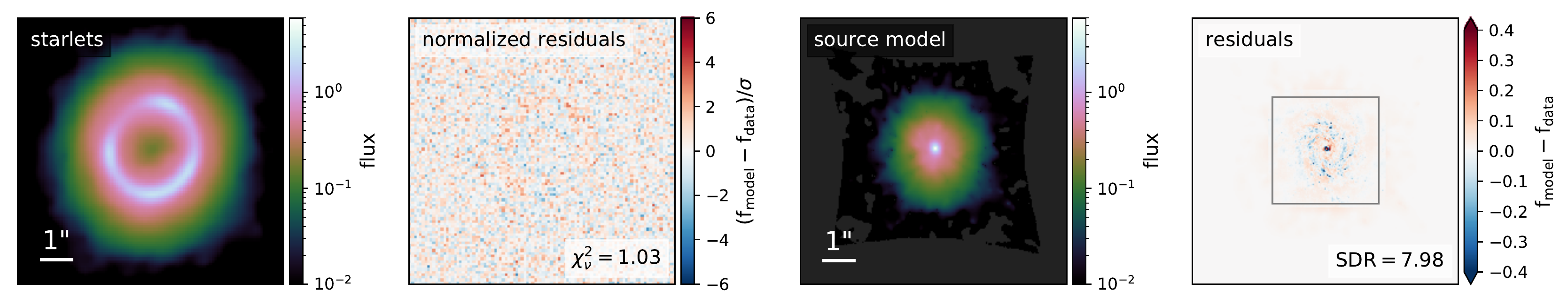}
    \includegraphics[width=\linewidth]{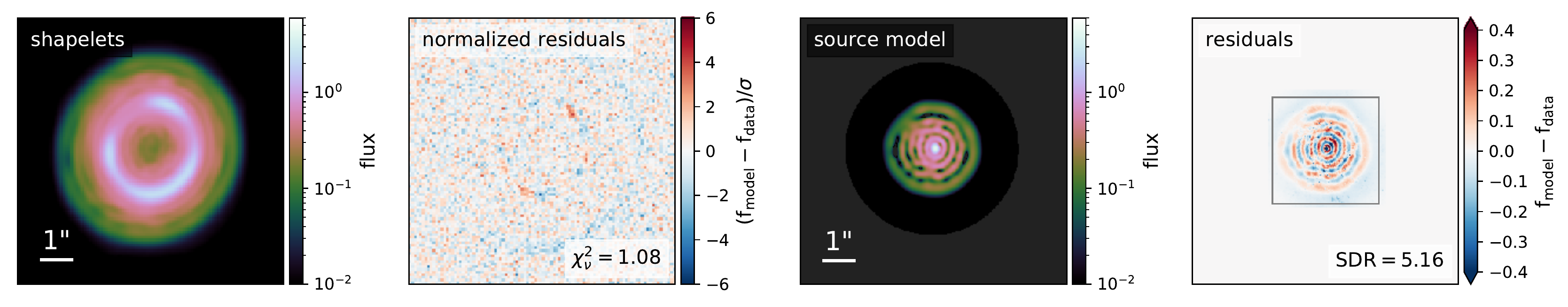}
    \includegraphics[width=\linewidth]{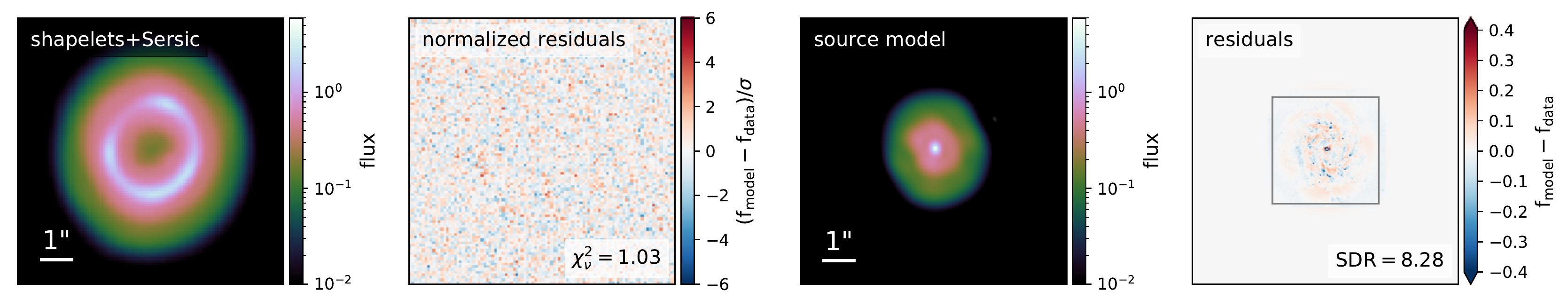}
    \caption{Source modelling of the simulated data with a single source galaxy (Fig.~\ref{fig:mock_HST}). \textbf{Top to bottom}: sparse model with starlets ($\pixsizeratio = 3$), analytical model with shapelets ($n_{\rm max}=18$), analytical model with shapelets+\sersic profile ($n_{\rm max}=8$). \textbf{Left to right}: image model, image normalised residuals and reduced chi-square, source model, source residuals and SDR. The SDR is computed only in the region indicated by the gray box, to avoid pixels with no flux to bias its value. All source reconstructions are shown at the supersampled resolution corresponding to the chosen \pixsizeratio. Dark gray areas in model panels correspond to non-positive pixel values (which can be negative with shapelets). \href{https://github.com/aymgal/SLITronomy-papers/blob/master/paper_I/mock_source_reconstruction.ipynb}{\faGithub}}
    \label{fig:source_recon_mock_HST_sgl}
\end{figure*}

\begin{figure*}
    \centering
    \includegraphics[width=\linewidth]{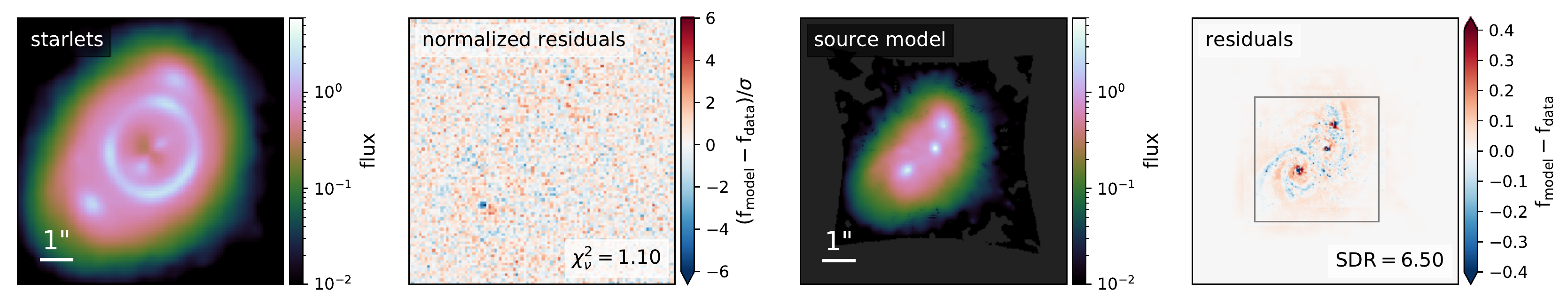}
    \includegraphics[width=\linewidth]{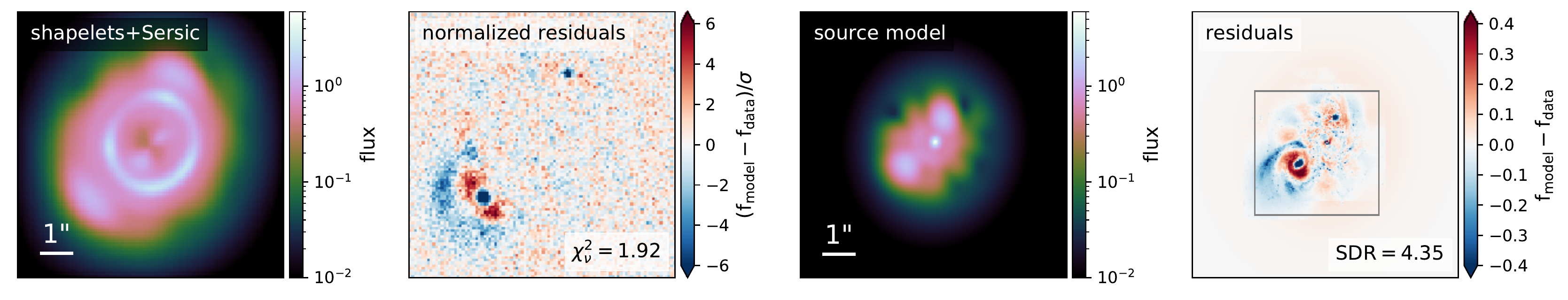}
    \includegraphics[width=\linewidth]{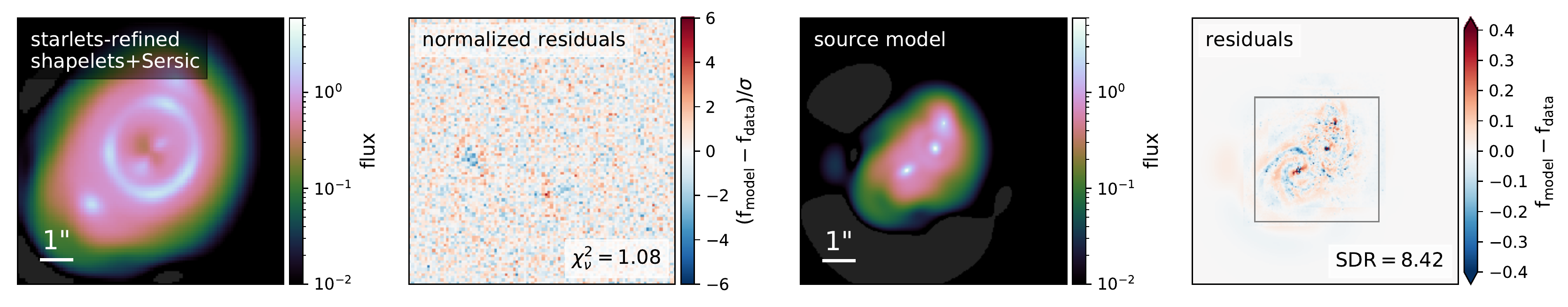}
    \caption{Source modelling of the simulated lensed galaxy group (bottom panel of Fig.~\ref{fig:mock_HST}). \textbf{Top to bottom}: sparse model with starlets ($\pixsizeratio=3$), analytical model shapelets+\sersic ($n_{\rm max}=8$), and starlets-refined shapelets+\sersic model ($n_{\rm max}=\{3, 3, 9\}$). \textbf{Left to right}: image model, image normalised residuals and reduced chi-square, source model, source residuals and SDR. The SDR is computed only in the region indicated by the gray box. All source reconstructions are shown at the supersampled resolution corresponding to the chosen \pixsizeratio. Dark gray areas in model panels correspond to non-positive pixel values (which can be negative with shapelets). We emphasise that the iterative refinement process, uses the model in first row to automatically setup the low-complexity model in second row, and further refine it until the reconstruction shown in last row (see Fig~\ref{fig:starlets_refinement} and Appendix~\ref{app:starlets_refinement}). \href{https://github.com/aymgal/SLITronomy-papers/blob/master/paper_I/mock_source_reconstruction.ipynb}{\faGithub}}
   \label{fig:source_recon_mock_HST_grp}
\end{figure*}

\begin{figure*}
    \centering
    \includegraphics[width=\linewidth]{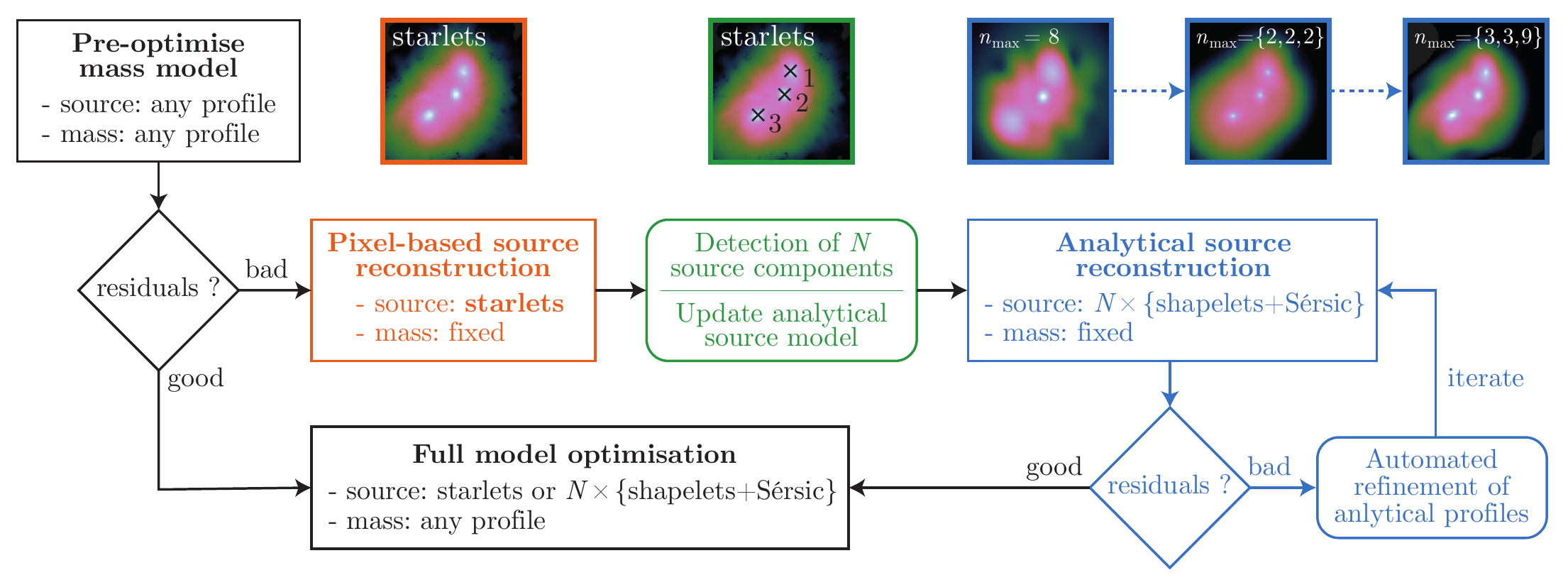}
    \caption{Workflow for using the starlet model as an automatic refinement of the analytical model. From a first estimation of the lens model, model residuals are assessed through visual inspection and/or goodness-of-fit statistics. If the current source model is insufficient, the lens model is fixed and the source is reconstructed using sparse optimisation with starlets. Potential sub-components of the source are located in source plane through special filtering followed by local maxima detection. A pair of profiles shapelets+\sersic is assigned to each sub-components, with low complexity (shapelet order $n_{\rm max}=2$). An iterative process is engaged for refining individual shapelet orders based on residuals in pixels corresponding to their sub-component (see Appendix \ref{app:starlets_refinement} for details). Once refined orders are sufficient, a full model optimisation can be performed.}
    \label{fig:starlets_refinement}
\end{figure*}

We demonstrate the source reconstruction capabilities of our starlet regularisation on the simulated data described above. As detailed in Sect. \ref{sec:slit}, the regularisation strength is defined by the detection threshold we wish to obtain in the source plane, in units of the noise. We set $\lambda'=3$ (i.e. $3\sigma$ significance for the source reconstruction) for all our models. We find that this permissive value leads to a good fit to the data for all applications in this work. We show in Appendix~\ref{app:effect_Js} that the number of scales has small impact on the source modelling at the level required for this work, but has substantial impact on the computation time. We fix the number of decomposition scales to $\Jscales{\source}=6$, giving reasonable computation time. We set the ratio of data pixel size to source pixel size to $\pixsizeratio=3$. As we focus on the source reconstruction, lens model parameters are held fixed to their input values.

We show in the first row of Figs.~\ref{fig:source_recon_mock_HST_sgl} and \ref{fig:source_recon_mock_HST_grp} results of our sparse reconstruction technique using starlets. We show the image model along with image plane normalised residuals, and the source model along with source plane residuals. The reduced chi-square values, $\chi^2_\nu$, are computed considering the number of (un-masked) pixels $i$ as the number of effective degrees of freedom $\nu$:
\begin{align}
    \label{eq:def_chi2nu}
    \chi_\nu^2 = \frac{1}{\nu} \sum_{i} \frac{\left(f_{\rm model,\,i} - f_{\rm data,\,i}\right)^2}{\sigma_i^2}\ ,
\end{align}
where $\sigma_i^2$ is noise variance of pixel $i$ (Eq.~\ref{eq:def_noise}).

In the source plane, we assess the quality of the reconstruction by computing the source distortion ratio \citepalias[SDR, as defined in][]{Joseph2019}:
\begin{align}
    \label{eq:def_sdr}
    {\rm SDR} = 10\log_{10}\left(\frac{\normtwo{\source_{\rm true}}}{\normtwo{\source-\source_{\rm true}}}\right)
\end{align}
where $\source_{\rm true}$ is the true light distribution of the source. With such a definition, the higher the SDR, the better the reconstruction.

Both image plane and source metrics are shown in Figs.~\ref{fig:source_recon_mock_HST_sgl} and \ref{fig:source_recon_mock_HST_grp}. The imaging data is modelled almost to the noise level, as one can expect from the assumed knowledge of the lens model, noise covariance and PSF. Despite small artifacts incorrectly filtered out by the starlet regularisation \citepalias[similar features were noticed in][]{Joseph2019}, the global shape of the reconstructed source galaxy is in good agreement with the truth. One can distinguish smaller scale features such as the most prominent spiral arms, even though the deconvoluton of the source (jointly performed by the algorithm) is limited by the fairly large FWHM of the F160W PSF.

It is worth pointing out that although we use parts of the same software to generate and to model the data, the actual algorithms used in each process are importantly different. The lensed sources were simulated using analytical ray-tracing from a spline-interpolated image with much higher resolution than that of the final data. In contrast, the source modelling was performed by a different approach, both when using starlets (via the lensing operator and bilinear interpolation) and shapelets (via analytical evaluation of basis functions). We therefore avoid the logical inconsistency of attempting to model the simulated data using the same algorithm that produced it. This also means that we cannot expect a perfect reconstruction, even assuming the true lens model.

\subsubsection{Comparison with analytical shapelets \label{sssec:hst_shapelets}}

We next perform the same exercise with the analytical, yet flexible, modelling method using shapelet basis functions \citep{Bernstein2002,Refregier2003,Birrer2015}. While shapelets can potentially produce biases for shape measurements \citep[e.g.][in the context of weak lensing]{Melchior2010}, they have been successfully employed in strong lens modelling. Shapelets are Gauss-Hermite polynomials that form a complete and orthonormal basis set. To ensure a tractable optimisation of shapelet coefficients, only a limited number $n_{\rm max}$ of polynomials are considered to model the light distribution. A shapelet basis set is thus defined by three free non-linear parameters aside the fixed $n_{\rm max}$: the two coordinates of the basis center and the reference scale $\beta$. The total number of basis functions (linear amplitudes) is $n_{\rm tot}=(n_{\rm max}+1)(n_{\rm max}+2)/2$. The polynomial order and the reference scale set the minimal and maximal scales that can be resolved $\beta_{\rm min} = \beta\left(1+n_{\rm max}\right)
^{-1/2}$ and $\beta_{\rm max} = \beta\left(1+n_{\rm max}\right)
^{1/2}$, respectively.

There are two ways of modelling the source with shapelet basis functions. They can on one hand be employed as a standalone light profile \citep[e.g.][]{Birrer2015,TagoreJackson2016,Birrer2017}. Alternatively, they can be considered as small-scale corrections to an underlying large-scale light profile, commonly taken to be an elliptical \sersic profile whose centroid coincides with the shapelets centroid \citep{Birrer2019,Shajib2019,Shajib2020}. For completeness we compare our reconstruction technique with both approaches, which we term ``shapelets'' and ``shapelets+\sersic'', respectively. We set the maximal polynomial order $n_{\rm max}$ as the lowest value required to obtain image plane residuals down to the noise, based on visual inspection.

We optimise the non-linear parameters of the shapelet basis (scale and basis center) using the Particle Swarm Optimiser \citep[PSO,][]{Kennedy2001} implemented in \lenstro, which is particularly suited for finding global minima in large parameter spaces\footnote{We also used the Nelder-Mead simplex algorithm (from the \textsc{scipy} library), but found that it quickly fails at exploring the parameter space once more complex light profiles such as shapelets+\sersic or large shapelet basis are being optimised, given the constraints of HST images.}.

We show in Fig.~\ref{fig:source_recon_mock_HST_sgl} both the ``shapelets'' and ``shapelets+\sersic'' reconstructions, for direct comparison with the ``starlets'' reconstruction. One notices that both analytical and pixel-based methods display artifacts in source plane, although not at the same level. While starlets only introduce low significance false detections due to noise propagation from image plane, shapelets introduce prominent concentric rings that attempt to compensate small scale features in the source structure, such as the cuspy bright core of the spiral source.

We emphasise that starlets are able to represent the large dynamic range of source features on their own, while shapelets are not. By adding a \sersic profile -- or any other smooth profile -- to the shapelet functions, one can split the modelling of small- and large-scale features. Indeed, the \sersic profile acts as a low-pass filter, with shapelet functions capturing the remaining high frequencies and requiring a lower maximal order compared to shapelets alone. The global shape of the source is retained with shapelets+\sersic, even though the boundaries of the light distribution are sharper than with starlets (the truth seems to lie in between). For small-scale features, starlets show more fidelity to the true underlying surface brightness. Such features, when revealed with shapelets, need to be interpreted with care, since they may not be real, but rather artifacts from their limited flexibility.

While the starlet model provides slightly better normalised residuals and a smoother, more realistic source, the resulting~$\chi^2_\nu$ is very close to the one obtained with the shapelets+\sersic model. To reach this value with the analytical model, additional constraints through the \sersic profile are required, which are not accounted for in the~$\chi^2_\nu$. We therefore believe our model performs better, but this is not well captured by the small difference in~$\chi^2_\nu$. In contrast to just the likelihood value, a metric based on the posterior value --- which has to be carefully designed so that it takes into account intrinsic differences between the methods --- should be preferred, but this is beyond the scope of this work.

Similar observations can be made for the case of a source composed of multiple complex components. For the shapelets+\sersic reconstruction, a single shapelet basis is not sufficient to obtain residuals down to the noise, requiring an $n_{\rm max}$ so large that it negatively affects the convergence efficiency of the optimisation and introduces significant artifacts in source plane (similar to the middle panel in Fig.~\ref{fig:source_recon_mock_HST_sgl}). Only after carefully optimising multiple shapelets+\sersic pairs at the location of each individual sub-component in source plane can the residuals be reduced to a satisfactory level. The initial location of these analytical profiles is crucial to prevent strong degeneracies from appearing during optimisation. If one is interested in modelling only a few strong lens systems, the initialisation can be set manually by the investigator, based on model residuals. For batch modelling of large datasets, however, this is not a viable solution.

To tackle this limitation, we design a workflow motivated by the advantages of both analytical and pixel-based reconstruction methods. While the former offers a considerable gain in speed -- mostly when solving for a large number of non-linear parameters from lens and source models jointly -- the latter does not require any choice of light profiles, nor a good set of initial parameter values for initialisation. Hence, analytical methods are better suited to rapidly explore the initial, potentially large, parameter space, and to provide a starting point for further refinement steps. These refinement steps can then be exclusively pixel-based or, as we propose here, use an intermediate solution between pixel-based and analytical. The goal of such a hybrid approach is to retain timing efficiency, while substantially reducing human choices by including pixel-based flexibility.

The proposed workflow can be summarised in four steps, illustrated in Fig.~\ref{fig:starlets_refinement}: based on a pre-optimised lens model obtained through a faster, fully-analytical modelling step, (1) the sparse starlet technique is used to obtain a high-resolution model of the source; (2) the number and location of each sub-component of the source are automatically detected on the starlet model; (3) individual pairs of shapelets+\sersic are assigned to sub-components and their centres properly initialised with low polynomial order (typically $n_{\rm max}=2$); (4) the polynomial order of each shapelet basis is iteratively refined based on image plane residuals that are specific to the individual sub-components. Finally, a full lens model optimisation can be performed, either with the updated analytical or pixel-based source model, depending on the specific goal and requirements (e.g. flexibility, computation time). We refer to Appendix \ref{app:starlets_refinement} for more details. We note that although the initial lens model estimation needs to be good enough to discard or detect multiple source components, the pixel-based starlet reconstruction can be used to quickly assess of the reliability of the lens model (in addition to model residuals).

The result of this automated refinement is shown in the bottom row of Fig.~\ref{fig:source_recon_mock_HST_grp}. The final source model consists of three shapelets+\sersic pairs with joint centroids and maximal polynomial orders of $\{3, 3, 9\}$. The higher order of the third shapelet basis is necessary to model the most prominent spiral arm. As expected, residuals are drastically improved by this automated procedure, reaching a similar level as the starlet reconstruction in image plane. In source plane, the SDR is slightly better than with the pixel-based approach, showing visually similar residuals when compared to the true source. The flexibility and simplicity of sparse modelling is directly used to improve the large set of analytical profiles, still driven by the imaging data itself. Some works have already attempted to minimise investigator time for batch modelling of strong lens systems \citep[e.g.][]{Nightingale2018,Shajib2019,Shajib2020slacs}. One can imagine the procedure developed here to be part of a larger automated modelling pipeline in the future.

\subsection{Source reconstruction of SLACS lenses \label{ssec:hst_SLACS}}

\begin{figure*}
    \centering
    \includegraphics[width=\linewidth]{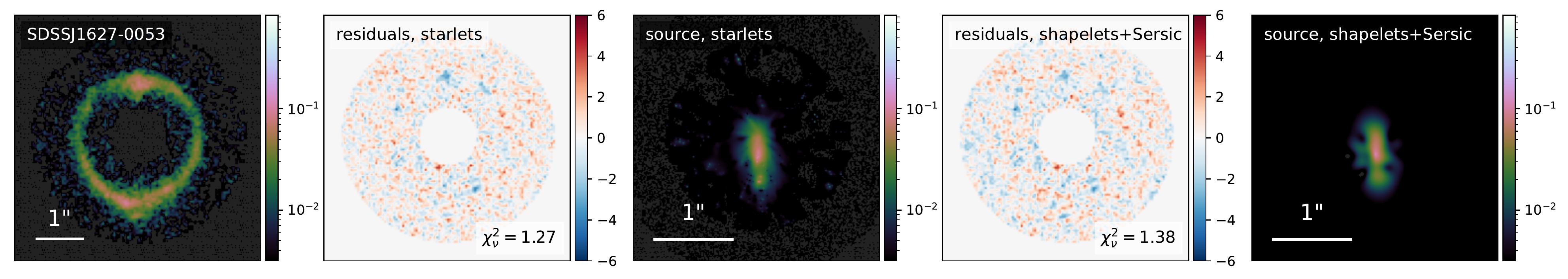}
    \includegraphics[width=\linewidth]{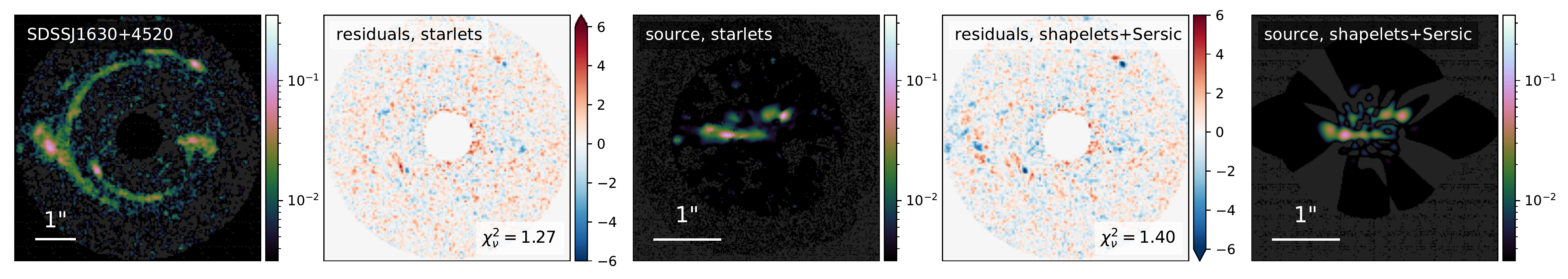}
    \includegraphics[width=\linewidth]{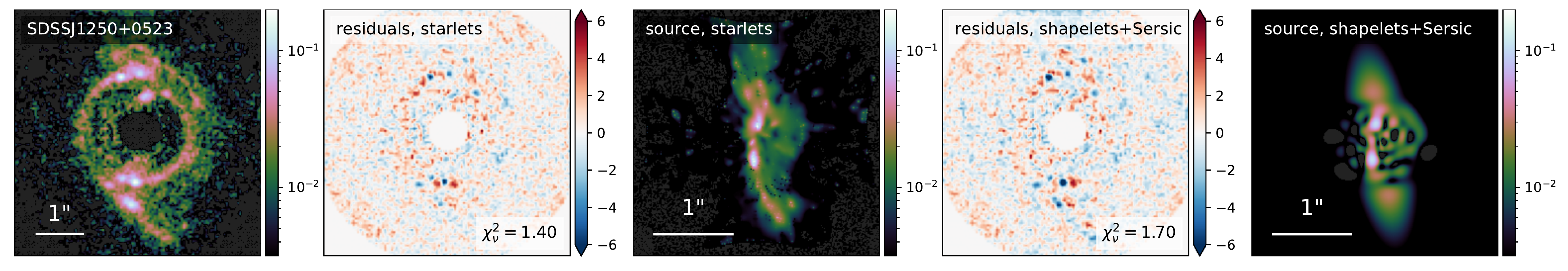}
    \includegraphics[width=\linewidth]{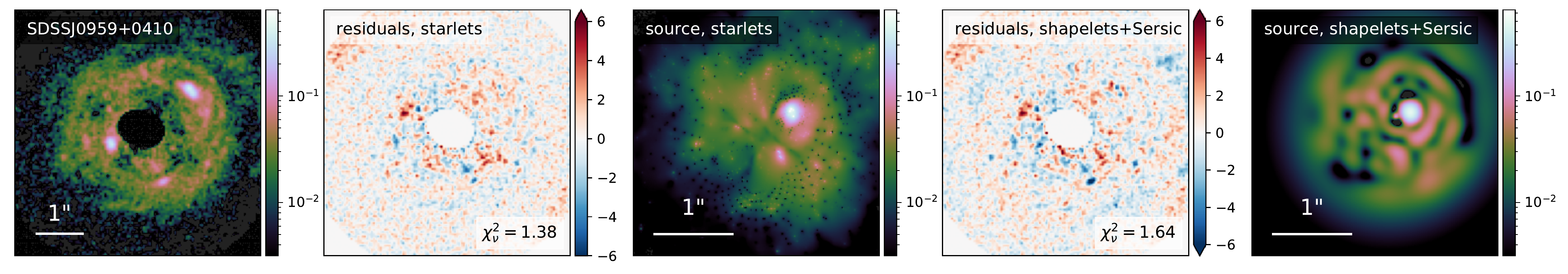}
    \caption{Source reconstruction of SLACS lenses, assuming analytical mass and lens light models from \cite{Shajib2020slacs}. Left to right: imaging data with lens-light subtracted, model residuals with shapelet model, shapelet source model, model residuals with starlet model, starlet source model ($\pixsizeratio=3$). The maximal polynomial order $n_{\rm max}$ used for shapelet models are, from top to bottom: 6, 15, 12, 15. We refer to the text for more details, specifically concerning the case of \zeroneufcinquanteneuf. Dark gray areas in model panels correspond to non-positive pixel values (which can be negative with shapelets). All source reconstructions are shown at the supersampled resolution corresponding to the chosen \pixsizeratio. Colour maps for flux and residuals retain their meaning from Fig.~\ref{fig:source_recon_mock_HST_grp}. \href{https://github.com/aymgal/SLITronomy-papers/blob/master/paper_I/SLACS_source_reconstruction.ipynb}{\faGithub}}
    \label{fig:slacs_fixed_mass}
\end{figure*}

We now apply \slitro to real data, focusing on a subset of lenses from the Sloan Lens Advanced Camera for Surveys sample \citep[SLACS,][]{Bolton2006}. This sample of 80 galaxy-galaxy lenses has been well studied and used, in combination with spectroscopy, to infer statistical properties of lens galaxies such as their dark matter and baryonic mass distributions, their location in the fundamental plane, as well as their selection functions \citep[e.g.][]{Gavazzi2007,Auger2010}. However, little focus has been given to the quality of the source reconstructions. It has also been shown that some properties of source galaxies can be recovered to sufficient precision without using pixel-level modelling \citep[e.g. for deriving AGN mass-luminosity relation,][]{Ding2020}. However, for certain applications that require insights on small-scale features in source galaxies, such as the study of star-forming regions \citep[e.g.][]{James2018} or galaxy morphologies, more flexible methods are required to reconstruct complex sources.

In our analysis, we assume that the lens mass model has been optimised beforehand. We use the recent work of \cite{Shajib2020slacs}, which uniformly modelled 23 of the SLACS lenses to put constraints on the mass profile and content of large elliptical galaxies. These models were also used as an external dataset in the hierarchical Bayesian analysis for time-delay cosmography introduced by \cite{Birrer2020}. Here we briefly summarise the modelling choices and fitting procedure used to optimise the lens model parameters (implemented in the \textsc{dolphin} package, a wrapper around \lenstro\footnote{\url{https://github.com/ajshajib/dolphin}}), and refer the reader to \cite{Shajib2020slacs} for more details. The mass profile of the lens is a power-law ellipsoidal mass distribution (PEMD), embedded in an external shear. The lens light (although masked for some systems) is modelled with a double elliptical \sersic profile. The source light is modelled with a shapelet function basis plus an elliptical \sersic profile. For an individual SLACS system, the adopted fitting recipe consists in a series of particle swarm optimisation (PSO) runs, each time altering specific degrees of freedom in the model: (1) the PEMD slope and external shear are fixed; (2) the flux from lensed arcs is masked for fitting of lens light; (3) all mass and light parameters for the deflector are fixed for fitting the source light, with fixed shapelet scale; (4) the PEMD parameters are relaxed for jointly optimising the source and mass profile parameters; (5) the shapelet scale is relaxed for further refinement of the source; (6) steps 2--5 are repeated with current best-fit values; (7) the PEMD slope and external shear are relaxed for a final MCMC sampling of all parameters together.

The modelling of \cite{Shajib2020slacs} has been performed with \lenstro, hence lens models are given in a format usable by our \slitro implementation with minimal pre-processing. Among the 23 lenses in the modelled sample, we first select one ``simple'' system, \seizevingtsept, that exhibits no particularly complex features (it is an almost perfect Einstein ring), and from which lens light can be removed properly. Three additional systems, intentionally selected for their higher complexity, are added to our subset of lenses, based on inspection of residuals from the analytical model and source reconstruction (shapelets+\sersic). They are \douzecinquante, \seizetrente and \zeroneufcinquanteneuf. We use HST data obtained with the ACS instrument in the F555W band for all four systems. We note that all of these systems have been considered to have good quality models by \cite{Shajib2020slacs}, although the main goal of their work was to obtain reliable lens model posterior distributions. We then take their best-fit lens model parameters\footnote{Private communication.} and apply our starlet reconstruction technique directly on the same imaging data cutouts, using the same PSF kernel as \cite{Shajib2020slacs} for the joint deconvolution. We use $\pixsizeratio=3$ for source plane resolution.

Figure~\ref{fig:slacs_fixed_mass} shows source reconstructions of the four systems. From left to right are the lens-subtracted imaging data, the residuals and shapelets+\sersic model of \cite{Shajib2020slacs}, followed by the residuals and model obtained with starlets. Reduced chi-square metrics are also indicated. Similarly to the simulated data of Sect.~\ref{ssec:hst_sims}, we notice that shapelet reconstructions for these systems display ring-like or wave-like features that do not likely represent the true source surface brightness. Additionally, the modelled light profile contains unphysical negative values, especially in the case of \seizetrente.

We observe a significant decrease in the residuals when using the sparse source reconstruction, compared to the original residuals obtained with the analytical model. While this may seem expected at first sight -- as the former model is much more flexible -- it should not be overlooked. Since we perform the pixel-based reconstruction at fixed mass in this case, it hints that lens model parameters that would be favored by the sparse reconstruction are compatible with those obtained from analytical optimisation. If this were not the case, residuals would, on the contrary, have been worse than previously obtained, because the sparse reconstruction would have favored another location in parameter space. Under the assumption that shapelet source reconstruction does not lead to biased lens model parameters \citep[which is supported by previous works][]{Birrer2015,TagoreJackson2016}, this is a reassuring result.

Specifically in the case of \zeroneufcinquanteneuf, we observe small-scale artifacts in the form of an array of low-amplitude spots that are clearly non-physical. They are a consequence of the pixel-based nature of our modelling technique, for which we identify three reasons. First, the source object -- likely to be two or more nearby galaxies -- extends over a region much larger than the caustics where the lens mapping tends to suffer more from the pixelated approximation. Second, the analytical lens light model is particularly inaccurate due to prominent overlapping of the lens and lensed arcs, in addition to a likely dust absorption lane (visible before lens light subtraction) that cannot be modelled well with smooth profiles only. Third, the mask defined to exclude pixels affected by the imperfect lens light subtraction causes an artificial discontinuity between the lensed arcs and the masked region, which is propagated to the source plane at the location of the artifacts. These three effects lead to inaccuracies in the noise estimation and cause source pixels at these locations to not be properly regularised when applying the sparsity constraint. We investigate these points further in Appendix~\ref{app:artifacts_origin}.

Pixel-based methods tend to be subject to over-fitting, which can artificially improve residuals. However, we argue this is not the case here, as one can easily identify features in the residuals that still persist after the starlet reconstruction of the source. Overall, no overfitting features clearly stand out from the residual maps. Our method's careful handling of noise levels, namely their propagation to starlet space and the subsequent application of sparsity constraints, all help to avoid overfitting.

\subsection{Optimisation of the analytical mass model} \label{ssec:mass_sampling_mock_HST}

\begin{figure}
    \centering
    \includegraphics[width=\linewidth]{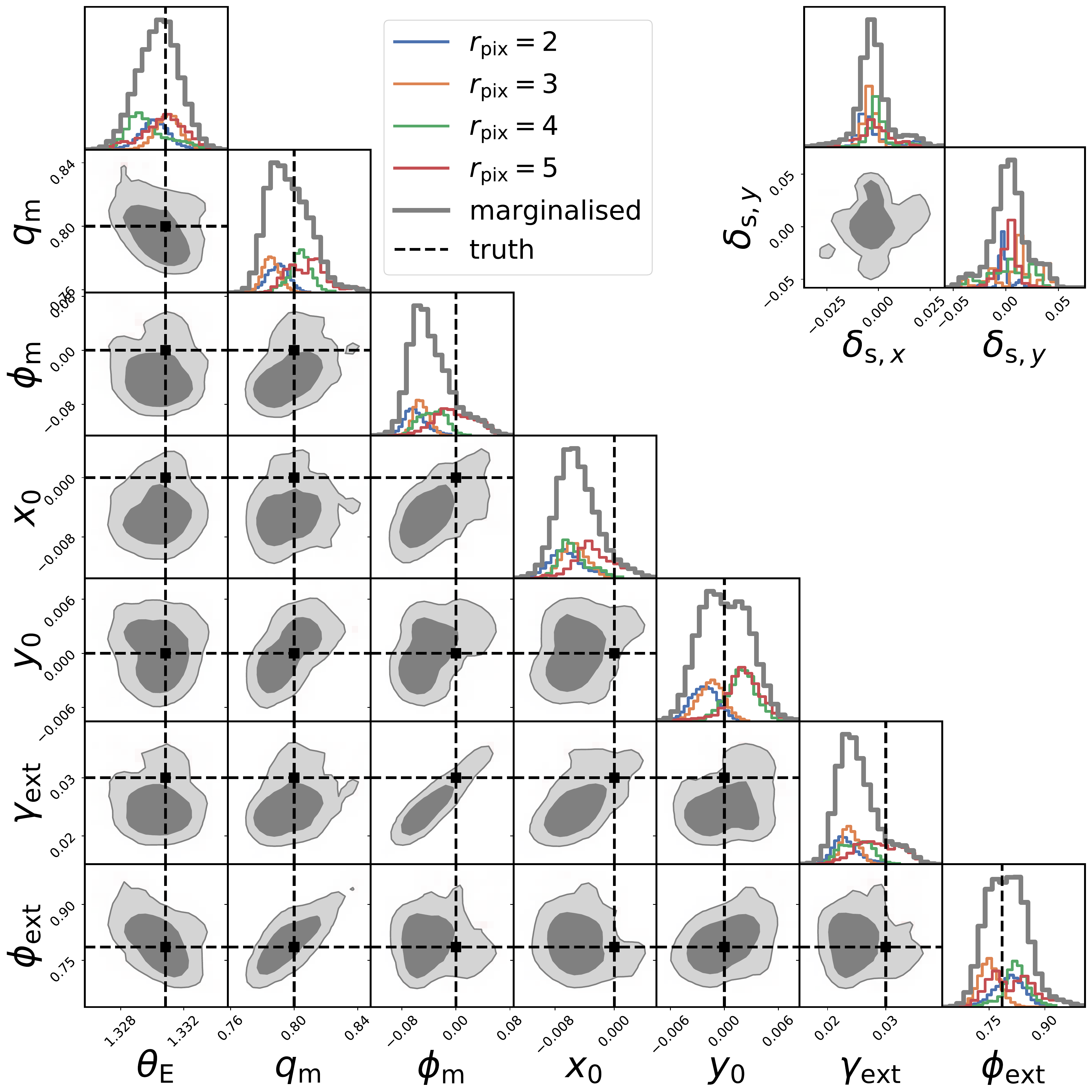}
    \caption{Posterior distributions for lens model (SIE + external shear) and source grid offset parameters (see Sect.~\ref{ssec:source_offset} for details), for the simulated HST data shown in Fig.~\ref{fig:mock_HST} (first row). We show individual posteriors for different values of \pixsizeratio (thin colored contours), and the marginalised posteriors (thick gray contours) have been obtained by combination with equal weights. Two-dimensional contours show 68\% and 95\% of the distribution volume, and true values are indicated by dashed lines.~\href{https://github.com/aymgal/SLITronomy-papers/blob/master/paper_I/mass_sampling/corner_plot_mass_sampling.ipynb}{\faGithub}}
    \label{app:fig:mass_sampling}
\end{figure}

Although our primary focus is on the problem of source reconstruction, estimating deflector mass parameters is also of interest for many strong lensing applications. As an example of how this can be done in our framework, we use MCMC\footnotetext{The MCMC is performed using \lenstro, which runs routines from the python package \textsc{emcee} (\url{https://github.com/dfm/emcee}).} to sample the lens model parameter space of our simulated HST data in the single source galaxy case (top row of Fig.~\ref{fig:mock_HST}). The resulting posterior distributions for SIE and external shear parameters are shown in Fig.~\ref{app:fig:mass_sampling}. We have checked the MCMC chains for convergence and verified that a sufficient number of samples has been discarded (we kept the last 35'000 samples each chain). By comparing the maximum \textit{a posteriori} values to the true input values, we observe that the source pixel size (controlled through \pixsizeratio) acts as the main source of systematic errors. This is consistent with what has been observed from other pixelated source reconstruction methods \citep[see e.g. Fig.~3 from][]{Suyu2013}.

In order to ensure robustness of the posterior and prevent underestimating error bars, we also marginalise over the different choices of \pixsizeratio. As there is no natural choice for this parameter, we apply equal weights to the individual distributions tp obtain marginalise posteriors. The resulting posteriors are all statistically consistent with the true parameter values. This demonstrates that our source regularisation method is suited to lens model parameter estimation as long as we properly account for the choice of source pixel size. We note that using a source pixel size equal to that of the data (i.e. $\pixsizeratio=1$) induces large pixelation effects and does not lead to an acceptable fit to the data (see Fig.~\ref{fig:lens_mapping}).

We have shown that a hybrid procedure of estimating the source on a pixelated grid while describing the lens mass through analytical functions does provide consistent results on simulated data. This can be extrapolated to real data only if real lens galaxies are indeed described accurately by simple analytical functions. However, if lens galaxies contain substantial complexity, such as non-elliptical mass distributions and dark substructure within the lens or along the line-of-sight, mass models must incorporate more flexibility to account for potential deviations from simple analytical distributions. Moreover, if the source model has significantly more flexibility than the lens model, part of mass complexity can be absorbed in the source reconstruction, leading to biased lens model parameters \citep{VernardosInPrep}.

\section{Data in the era of ELTs \label{sec:elt_data}}

\begin{figure}
    \centering
    \includegraphics[width=0.8\linewidth]{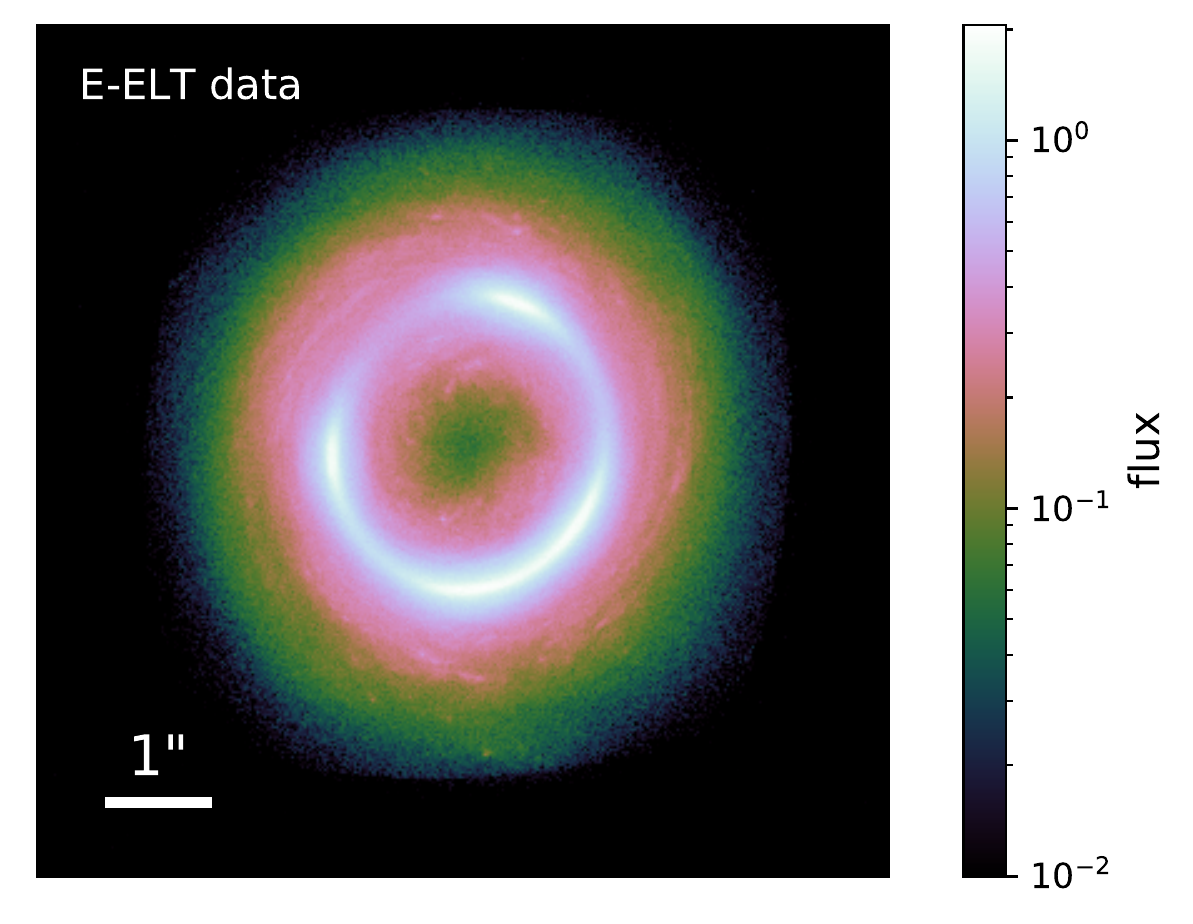}
    \caption{Simulated E-ELT data. The lens light is assumed to be perfectly subtracted. The seeing conditions and noise covariance are based on simulated observations in the $H$-band of the MICADO/MOARY instrument. The source galaxy is the same as the one in the top right panel of Fig.~\ref{fig:mock_HST}.}
    \label{fig:mock_ELT}
\end{figure}

\begin{figure*}
    \centering
    \includegraphics[width=\linewidth]{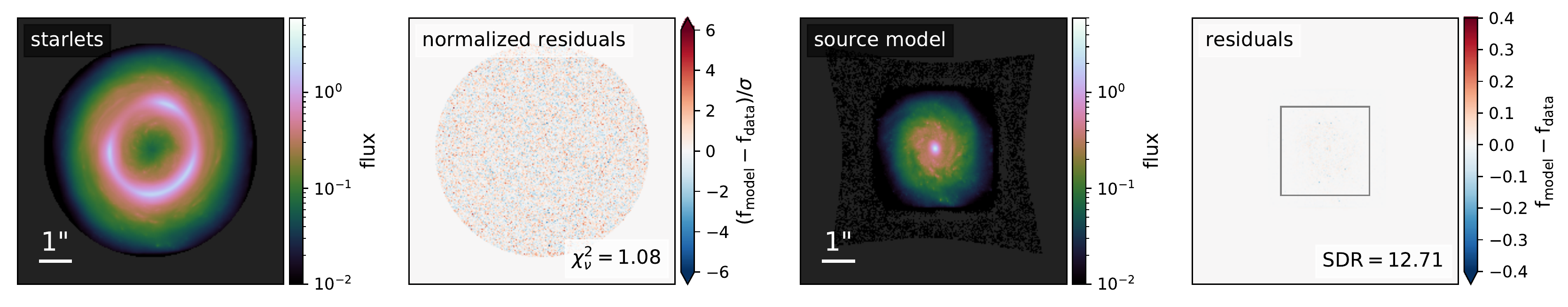}
    \includegraphics[width=\linewidth]{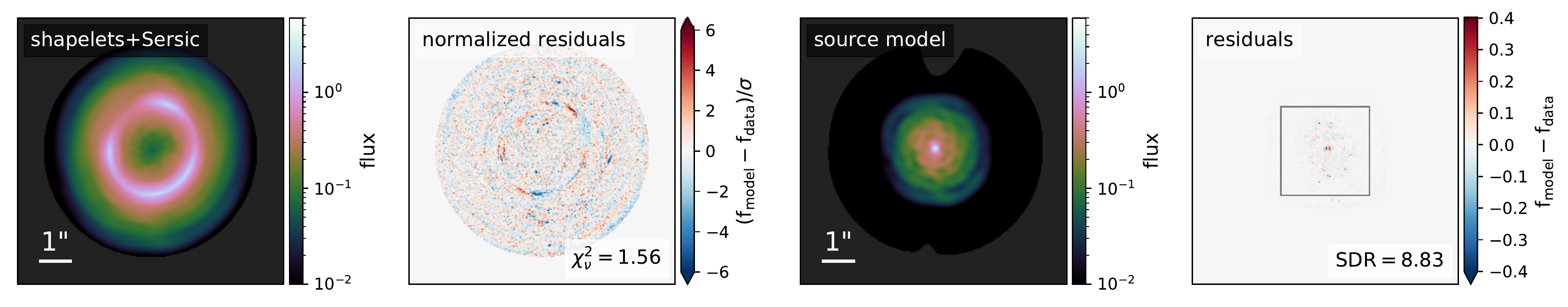}
    \caption{Source modelling of the simulated E-ELT data of Fig.~\ref{fig:mock_ELT}. \textbf{Top to bottom}: sparse model with starlets ($\pixsizeratio = 2$), analytical model with shapelets ($n_{\rm max}=20$). \textbf{Left to right}: image model, image-normalised residuals and reduced chi-square, source model, source residuals and SDR. The SDR is computed only in the region indicated by the gray box, to avoid pixels with no flux to bias its value. \href{https://github.com/aymgal/SLITronomy-papers/blob/master/paper_I/mock_source_reconstruction.ipynb}{\faGithub}}
    \label{fig:source_recon_mock_ELT}
\end{figure*}

\begin{figure*}
    \centering
    \includegraphics[width=\linewidth]{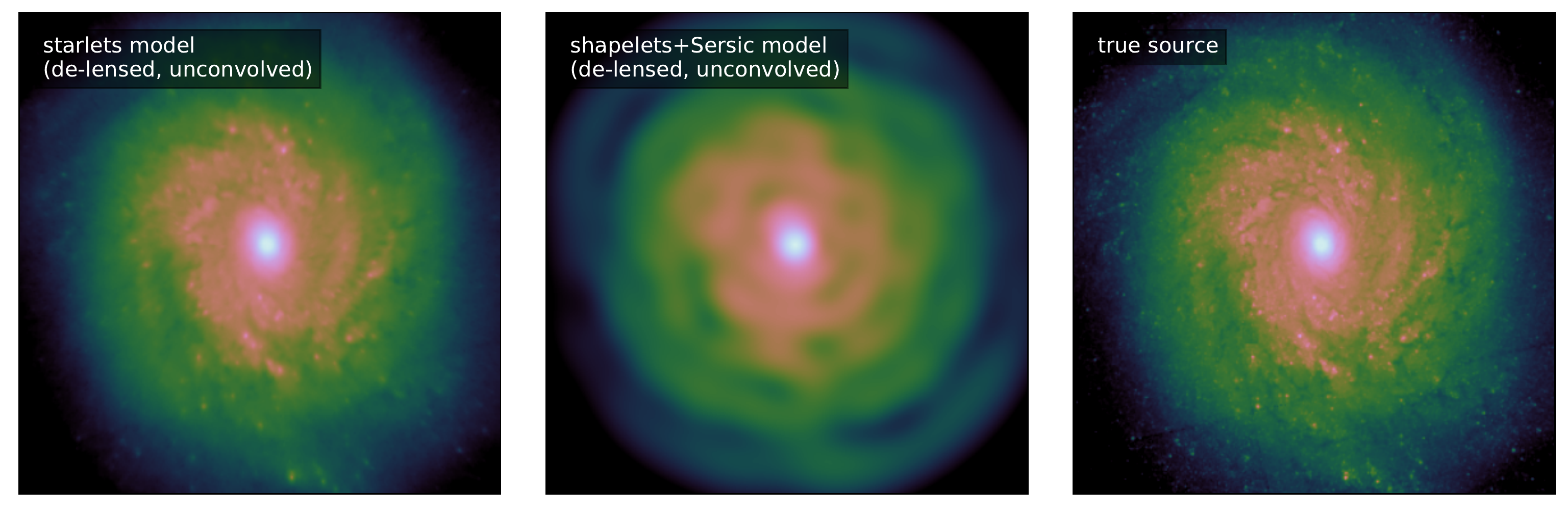}
    \caption{Zoom of the source reconstructions shown in Fig.~\ref{fig:source_recon_mock_ELT}, from the modelling of simulated E-ELT data}. From left to right: source reconstruction with starlets ($\pixsizeratio=2$), source reconstruction with shapelets+\sersic profile ($n_{\rm max}=20$), true source. We note that images are $2'000$ pixels on-a-side. The colour scaling is the same across to the panels. All source reconstructions are shown at the supersampled resolution corresponding to the chosen \pixsizeratio. \href{https://github.com/aymgal/SLITronomy-papers/blob/master/paper_I/mock_source_reconstruction.ipynb}{\faGithub}
    \label{fig:source_recon_mock_ELT_zoom}
\end{figure*}

Current observing facilities, such as HST and ground-based instruments with adaptive optics (e.g. VLT and Keck), allow us to constrain many galaxy properties due to their high-resolution imaging. Future thirty-meter-class telescopes will push image quality even higher, and our modelling techniques must be robust enough to accommodate such improved data. In this section, we simulate strong lensing images from a thirty-meter-class telescope and, similar to Sect.
~\ref{sec:hst_data}, we compare our sparse source reconstruction technique to an analytical strategy.

\subsection{Simulated E-ELT data}

Data such as those to be obtained with the future 39m European Extremely Large Telescope (E-ELT) are an obvious target for \slitro. The E-ELT will be equipped with, among other instruments, the near-infrared Imaging Camera for Deep Observations \citep[MICADO,][]{Davies2016}. It is designed to observe at the diffraction limit, either in Single Conjugate Adaptive Optics (SCAO) mode, or in Multi-conjugate Adaptive Optics mode (MCAO) powered by the MAORY facility \citep{Diolaiti2020}. This instrument is representative of the future generation of ultra-sharp imagers mounted on 30m class telescopes.

In the following we use the same high-resolution source galaxy as in Sect. \ref{sssec:mock_hst} (top right panel of Fig.~\ref{fig:mock_HST}), and the same lensing configuration (see Table~\ref{app:tab:sim_params}). We choose to render the image in the $H$ band of the MICADO imager, as it is the closest to the WCF3/F160W band of HST. The pixel scale is 4 mas, or twenty times smaller than the HST pixel size in the corresponding filter. We use the Exposure Time Calculator for the E-ELT imager to obtain typical settings for the MICADO/MAORY instrument in the $H$ band. We found that a single exposure of 1'200 seconds corresponds to a typical setting for the type of object we are interested in. After ray-tracing, the seeing is simulated by convolving the image with the $H$-band MCAO PSF provided by the MAORY consortium through the package \textsc{SimCADO}\footnote{\url{https://github.com/astronomyk/SimCADO}} \citep{Leschinski2016}. Finally, a mix of Gaussian and Poisson noise is added, consistently with the chosen setup. All simulation settings are summarised in Table~\ref{app:tab:sim_settings}. The lens and source properties are identical to the HST simulation of Sect.~\ref{ssec:hst_sims}, and are summarised in Table~\ref{app:tab:sim_params}.

We show in Fig.~\ref{fig:mock_ELT} the simulated E-ELT strong lens system. The final square cutout is $2'000\times2'000$ pixels, which is a challenging size for many of the modelling tools currently in use. This is true not only in terms of computation time, but also in terms of numerical pixel-level accuracy. One could reduce the effective data size by extracting smaller cutouts from the original image and modelling each of them individually along with a joint mass model. However, modelling the entire system at once is simpler, requires potentially less human interaction, and is expected to give much more constraining power as every pixel will contribute to the model. It also avoids the edge effects inherent to some lens inversion codes. 

\subsection{Source reconstruction with E-ELT data}

We follow Sect. \ref{sec:hst_data} and apply the \slitro algorithm to our simulated E-ELT data. We set the pixel scale of the source to half that of the data pixel, that is $\pixsizeratio=2$, which differs from the HST case. With E-ELT data, we find that the resolution is already high enough that going to smaller scales only increases the computation time considerably without improving the quality of the source reconstruction. This is due to the smaller data pixel size combined with the E-ELT PSF that is much better sampled than the HST PSF. We also exclude a large fraction of pixels that contain no flux from the lensed source by defining a circular mask of 6.4 arcsec in diameter, centred on the deflector. This is to avoid the large number of pixels that contain no flux from biasing the reduced chi-squared metric by artificially driving it to low values.

We show source reconstructions in Fig.~\ref{fig:source_recon_mock_ELT}. In the first row is the starlet model. In the image plane the reconstruction is down to the noise, although a few isolated pixels (not visible on the plot) appear above $6\sigma$ level. The resolution and signal-to-noise of the E-ELT data, coupled with the multi-resolution property of our regularisation method, allows us to model the source galaxy at high fidelity. We are able to recover most of the features at all spatial scales simultaneously, which emphasizes the advantage of using a genuine multi-scale reconstruction technique. As a result, small star-forming regions, larger-scale spiral arms and bulges, and even larger-scale luminous halos are equally well reconstructed.

Unlike with HST data, we expect an analytical model composed of shapelet basis functions superimposed on a smooth \sersic profile to show limitations in terms of source reconstruction, because shapelets cannot capture a broad range of spatial scales simultaneously. In the second row of Fig.~\ref{fig:source_recon_mock_ELT} we show such a reconstruction with a shapelets+\sersic model obtained with $n_{\rm max}=20$. Even with very high number of shapelet functions ($n_{\rm tot}=231$), the source cannot be modelled to the level allowed by the imaging data, both in the image and source planes.

Because of the high number of pixels in the source models, it can be difficult to visualise details. For this reason we show in Fig.~\ref{fig:source_recon_mock_ELT_zoom} a zoomed-in view of both reconstructions, along with the ground truth. The two reconstructions differ considerably, and we observe how the limited number of shapelet basis functions affects the minimal and maximal scales that can be recovered (which are fixed by the pair of parameters $\{n_{\rm max},\,\beta\}$). Star-forming regions and smaller spiral arms cannot be modelled, nor the larger-scale halo (despite the \sersic profile included). However, comparing the starlet reconstruction to the true source shows the level of detail that can be recovered thanks to sparsity and starlet constraints.

It is interesting to compare the computation time required by both methods, in this particular setting. In order to limit the number of calls to the loss function, we use the Nelder-Mead simplex minimiser to optimise the analytical model (with maximum number of iterations fixed to 50). While PSO is more efficient and better at finding the global minimum, it runs longer on average\footnotetext{We note that in \lenstro, other non-linear (e.g. lens model) are optimised jointly with shapelet parameters, so the computation time does not scale linearly in such a case.}. By using the former, one therefore gets an approximate lower bound on the computation time. The \slitro algorithm takes about 7 minutes to perform the reconstruction with $\pixsizeratio=2$, whereas the analytical reconstruction requires about 2.8 hours to optimise all shapelet basis functions and \sersic parameters. As the number of shapelet functions in the basis scales as~$\mathcal{O}(n_{\rm max}^2)$, further increasing the maximal polynomial order drastically increases computation time.

We have noticed that going to a higher number of analytical basis functions -- in combination to such a large number of data pixels -- implies to allocate and invert a very large matrix, that caused memory issues. This limitation can be partially addressed by performing the source reconstruction from a down-scaled version of the original data. This trade-off between data resolution and number of basis functions in the model constitutes an alternative choice to the standard shapelets+\sersic model discussed above. We also note that for a large number of basis functions, the computation time of the linear inversion step starts to dominate over the optimisation of non-linear parameters, such that in practice, suitable estimates of non-linear parameters are required.

While analytical methods have so far been efficient and accurate enough for many strong-lensing applications, they also have limitations. Where galaxies (lens or source) display small-scale and complex features, as can already be seen in current imaging data and will become the standard with upcoming thirty-meter-class telescopes, these techniques lack the flexibility to provide accurate source reconstructions. Moreover, their need for fine-tuning at the level of individual systems (in order to, for example, accurately recover sub-components of the lens) means increasingly more human time spent on modelling. Our model-independent sparse reconstruction technique alleviates these problems by being both highly flexible and efficient enough to handle high-resolution and morphologically complex data.

\section{Conclusion \label{sec:conclusion}}

In this work we have introduced a novel implementation of the Sparse Linear Inversion Technique (SLIT) of \citetalias{Joseph2019}, improved in many aspects. Our \slitro plugin is fully compatible with the modelling software \lenstro \citep{Birrer2018lenstro} and adds efficient pixel-based reconstruction capabilities relying on sparsity and wavelets to be used jointly with analytical techniques. We demonstrated the quality of source reconstruction on various cases using mock HST data, real strong lenses from the SLACS sample, and simulated E-ELT imaging data. We can summarise this paper as follows:

\begin{itemize}
    \customitem We use starlets, a specific type of wavelets, to represent the surface brightness of galaxies. Starlets are well suited to model complex sources without any need to choose an arbitrary number of analytical profiles to be stacked at arbitrary positions. No human intervention is required with wavelets, which is a genuine multi-scale decomposition of the data.
    \customitem The multi-scale property of the starlet transform allows us to use smaller source pixels without a large decrease in the quality of the source reconstruction. Sparse regularisation enforces the trade-off between the reconstruction of high frequency signals in the source and noise removal.
    \customitem For future high-resolution observations, like those obtained with thirty-meter-class telescopes, pixel-based methods may be the only viable option, as illustrated with simulated but realistic E-ELT data.
    \customitem Pixel-based methods may be limited by their higher computation time compared to analytical methods, as long as the number of pixels used to represent the source remains modest, like with HST data. However, with increasing numbers of pixels per resolution element, as is the case with E-ELT data, the trend is in fact inverted, and pixel-based methods such as \slitro become faster than analytical methods.
\end{itemize}
    
We see the present work as a first step toward building a full wavelet-based strong lensing method where all the elements to be modelled are described with wavelets. The current \slitro implementation allows us to optimise for the deflector mass within the modelling software \lenstro, jointly reconstructing the source using starlets. However, this implementation is currently limited by the use of analytical lens mass profiles, potentially lacking of flexibility to capture complex mass distributions such as large scale twists in the mass profile. Sparsity, combined with wavelets, is a powerful tool to reconstruct mass distributions in a model-independent way. It has been applied successfully in the context of weak lensing. Our future work in strong lensing will be focused on applying sparsity constraints for the deflector mass and light, hence leading to similar levels of flexibility in the modelling of both the source light and the lens mass and light. In doing so, it will be crucial to link the lens mass reconstruction to its kinematics, as the latter plays a vital role in alleviating the degeneracies inherent to the lensing data alone. Consequently, a full sparse modelling tool should also support the description of 2D or even 3D stellar kinematics in any complex potential well.

\begin{acknowledgements}
AG would like to thank the referee for their comments and suggestions that improved the content of the paper. AG warmly thanks S.~Birrer for useful feedback related to the integration of \slitro into the \lenstro interface, and feedback and ideas on the paper. AG thanks G.~Vernardos and M.~Millon for insightful discussions. AG thanks X.~Ding for providing the pre-processed HST images used for simulated source galaxies \citep[used in][]{Ding2020_tdlmc}. AG thanks A.J.~Shajib for providing the lens models for SLACS lenses \citep[see][]{Shajib2020slacs}. This program is supported by the Swiss National Science Foundation (SNSF) and by the European Research Council (ERC) under the European Union’s Horizon 2020 research and innovation program (COSMICLENS: grant agreement No 787886). This research made use of \lenstro \citep{Birrer2015,Birrer2018lenstro}, \textsc{SciPy} \citep{Virtanen2020scipy}, \textsc{NumPy} \citep{Oliphant2006numpy,VanDerWalt2011numpy}, \textsc{Matplotlib} \citep{Hunter2007matplotlib}, \textsc{scikit-image} \citep{VanDerWalt2014scikitimage}, \textsc{Astropy} \citep{astropy2013,astropy2018} and \textsc{Jupyter} \citep{Kluyver2016jupyter}.
\end{acknowledgements}


\bibliographystyle{aa}
\bibliography{biblio}


\begin{appendix}

\section{Properties of simulated HST and E-ELT data}

In this section we summarise all fixed parameters that define the simulations used for source reconstructions in Sects. \ref{sec:hst_data} and \ref{sec:elt_data}. They can be separated into two categories. Instrumental settings that control the resolution, seeing and signal-to-noise of the data are shown in Table~\ref{app:tab:sim_settings}. Model parameters including redshifts, mass distributions of the deflector and positions of the source galaxy(-ies) are specified in Table~\ref{app:tab:sim_params}.

HST instrumental settings are inspired by typical near-infrared imaging data required for time delay cosmography after drizzling. E-ELT data settings are inspired by \citet[][although for the Thirty Meter Telescope]{Meng2015} and \citet[][limiting magnitudes]{Deep2011}, and the E-ELT Exposure Time Calculator\footnote{\url{http://www.eso.org/observing/etc/bin/gen/form?INS.NAME=ELT+INS.MODE=swimaging}}.

Lens model parameters were picked to be fairly representative of an average strong lens system. The mass distribution is described by the singular isothermal ellipsoid (SIE), sustained in a moderate external shear field, and slightly misaligned with the SIE orientation.

We use the \texttt{SimulationAPI} module of \lenstro to generate realistic imaging data based on instrumental and model parameters above.

\begin{table}[htbp!]
\caption{Instrumental and observational settings used for HST and E-ELT simulated data. \href{https://github.com/aymgal/SLITronomy-papers/blob/master/paper_I/fixed_param.py}{\faGithub}} \label{app:tab:sim_settings}
\renewcommand{\arraystretch}{1.2}
\centering
{\small
\begin{tabular}{lcc}
\hline
& \textbf{HST} & \textbf{E-ELT} \\
\hline\hline
Instrument/filter & WCF3/F160W & MICADO/H-band \\
Pixel size [mas] & 80 & 4 \\ 
Single-exposure time [s] & 500 & 1'200 \\ 
Number of exposures & 4 & 1 \\ 
Zero-point [mag] & 27 & 34 \\ 
Sky brightness [mag/arcsec$^2$] & / & 25 \\ 
Read noise [e$^-$] & / & 3 \\ 
CCD gain [e$^-$/ADU] & 2.5 & 1 \\
Background noise [e$^-$] & 0.05 & / \\
PSF kernel (simulated) & \textsc{TinyTim} & \textsc{SimCADO} \\
\hline
\end{tabular}
}
\end{table}

\begin{table}[htbp!]
\caption{Model choices and parameters used for both HST and E-ELT mock data simulations. The coordinates are oriented such as North is up and East is right. The orientation is zero when aligned with the horizontal axis and increases anti-clockwise. Images of the source galaxies were obtained from the HST archival database. ``Source 1'' is the main source component in all simulations in this paper, whereas ``Source 2'' and ``Source 3'' are only used as member of the simulated galaxy group of Fig. \ref{fig:source_recon_mock_HST_grp}. \href{https://github.com/aymgal/SLITronomy-papers/blob/master/paper_I/fixed_param.py}{\faGithub}} \label{app:tab:sim_params}
\renewcommand{\arraystretch}{1.2}
\centering
{\small
\begin{tabular}{lc}
\hline
\textbf{Deflector}, $z_{\rm d}=0.3$ & \\
\hline\hline
\ \ \ Singular isothermal ellipsoid (SIE) & \\
\ \ \ \ \ \ Velocity dispersion, $\sigma_{\rm v}$ & 260 km/s \\
\ \ \ \ \ \ Ellipticity, $q_{\rm m}$ & 0.8 \\
\ \ \ \ \ \ Orientation, $\phi_{\rm m}$ & 0 \\
\ \ \ \ \ \ Position, (${\rm RA}=x_0$, ${\rm Dec}=y_0$) [arcsec] & (0, 0) \\
\hline
\ \ \ External shear & \\
\ \ \ \ \ \ Strength, $\gamma_{\rm ext}$ & 0.03 \\
\ \ \ \ \ \ Orientation, $\phi_{\rm ext}$ & $\pi/4$ \\
\hline
\textbf{Source}, $z_{\rm s}=1.2$ & \\
\hline\hline
\ \ \ Source 1: NGC3982 & \\
\ \ \ \ \ \ Position, (RA, Dec) [arcsec] & (0.05, 0.05) \\
\ \ \ \ \ \ Magnitude & 22 \\
\hline
\ \ \ Source 2: ESO498G5 & \\
\ \ \ \ \ \ Position, (RA, Dec) [arcsec] & (-0.8, -0.6) \\
\ \ \ \ \ \ Magnitude & 22 \\
\hline
\ \ \ Source 3: NGC3259 & \\
\ \ \ \ \ \ Position, (RA, Dec) [arcsec] & (0.3, 0.8) \\
\ \ \ \ \ \ Magnitude & 23 \\
\hline
\end{tabular}
}
\end{table}

\section{Hybrid approach for automated modelling \label{app:starlets_refinement}}

In Sect. \ref{sssec:hst_shapelets} we compare source reconstructions obtained with starlet and shapelet models on simulated HST data. For a complex source, the flexibility of a pixel-based approach outperformed the default analytical profile. However, analytical modelling has the advantage of being faster, and thus is better suited to batch automated modelling. Furthermore, for lower resolution large datasets like future surveys will deliver, the quality of analytical source reconstructions is often sufficient for reliable population-level lens model optimisation.

For these reasons, we propose a workflow using both pixel-based and analytical methods to refine in an automated manner an initially poor source model. Figure \ref{fig:starlets_refinement} illustrates the decision flow. Here we give a description of each step, assuming an approximation of the lens model has been previously obtained, for example through optimisation with a simple source profile:
\begin{enumerate}
    \item if current model residuals display features that are likely to come from the limiting source profile (e.g. after visual inspection or chi-square) we perform a pixel-based source reconstruction using sparse modelling;
    \item the reconstructed map of source surface brightness is smoothed using a Laplacian-of-Gaussian filter, to enhance individual source sub-components. A local maxima detection algorithm is applied on the fltered map to find the location of each sub-component;
    \item an analytical model is initialised as one joint shapelets+\sersic pair centered on each sub-component~$i$. Shapelet maximal orders are set to a low $n_{{\rm max},\, i}=2$;
    \item the source is re-optimised, and the quality of image plane residuals at the location of each lensed sub-component are used to iteratively refine the maximal order of each shapelet basis independently;
    \item once localised residuals are below a given threshold, a final full optimisation (mass+source) can be performed.
\end{enumerate}
The step refining individual shapelet polynomial orders requires establishing a suitable update rule to allow a smooth increase in model complexity without introducing too many degrees of freedom that may not be supported by the data. We found quite successful the following update rule: given a reduced $\chi^2_{\nu,\,i}$ based on residuals restricted to the location of the lensed sub-component $i$, the shapelet order is updated according to
\begin{align}
\label{app:eq:update_rule}
    n_{{\rm max},\,i}^{\rm new} = 
    \left\{
	\begin{array}{ll}
	    n_{{\rm max},\,i}^{\rm old} + \max{\left(1,\, 10\times\left\lfloor\chi^2_{\nu,\,i} - 1.1\right\rfloor\right)}\ \ \ {\rm if}\ \chi^2_{\nu,\,i} > 1.1\ ,
		\\
		n_{{\rm max},\,i}^{\rm old}\ \ \ {\rm if}\ \chi^2_{\nu,\,i} \leq 1.1\ .
	\end{array}
    \right.
\end{align}
Additionally, we set an overall maximal value of $n_{\rm max}=14$ to prevent the procedure above from leading to an intractable number of basis functions.

We see the procedure described in Fig.~\ref{fig:starlets_refinement} as a potential component of a larger pipeline for automatised modelling of strong lens systems. Various pipelines are currently in development based on recent works like the uniform modelling strategies proposed in \cite{Shajib2019,Shajib2020slacs}, the automatic differentiable pipeline of \cite{Chianese2020}, and workflows developed by the teams that took part in the Time Delay Lens Modelling Challenge \citep{Ding2020_tdlmc}.

\section{Threshold refinement for sparsity constraint \label{app:reweighting}}

As shown in Eq.~\ref{eq:slit_problem} we enforce sparsity on reconstructed surface brightness maps by minimising their $\ell_1$ norm in starlet space. Sparsity can also be enforced through the $\ell_0$ norm instead, but it is not convex like the $\ell_1$ norm and tends to produce in practice more artifacts in the presence of noise. However, enforcing minimal $\ell_1$ norm in the presence of noise has a greater dependence on the amplitude of each pixel, which can potentially bias the reconstruction. We employ a similar solution as proposed by \cite{Candes2007} by iteratively correcting for this bias through additional ``reweighting'' steps. Once the solution of the original unweighted problem has been found ($\Wweights{\mat{x}}^0= 1$), the minimisation is performed a second time with threshold per pixel per starlet scale updated according to the following rule, for a given iteration~$n$
\begin{align}
    \label{app:eq:reweighting}
    \Wweights{\mat{x}}^{n} = 
    \left\{
	\begin{array}{l}
        1 \ \ {\rm if}\ n=0 \, , \\
        \dfrac{1}{ 1 + \exp\!\left[10 \times \left(\starletsop\mat{x}^{n-1} - \lambda\right)\right] } \ \ {\rm if}\ n>0 \, ,
    \end{array}
    \right.
\end{align}
where $\mat{x}^{n-1}$ stands either for the source or lens light at the previous iteration, and $\lambda$ is the regularisation strength in units of noise per pixel in starlet space, $\lambda\equiv\lambda'\sigma_\starletsinvop$, after proper propagation through appropriate operators of Eq.~\ref{eq:slit_model}. In practice, a single additional optimisation loop with the weights updated according to the above equation already improve the source reconstruction significantly, as better residuals and lower chi-squared are obtained. Analogous to \citetalias{Joseph2019}, we found that the update rule of Eq.~\ref{app:eq:reweighting} is suitable for our lens modelling application. The specific form of the reweighting scheme is not unique, however it has to decrease the effective threshold for specific starlet coefficients that are significant enough, from previous iterations, relative to the noise.

\section{Artifacts in source model of \zeroneufcinquanteneuf \label{app:artifacts_origin}}

\begin{figure}
    \centering
    \includegraphics[width=\linewidth]{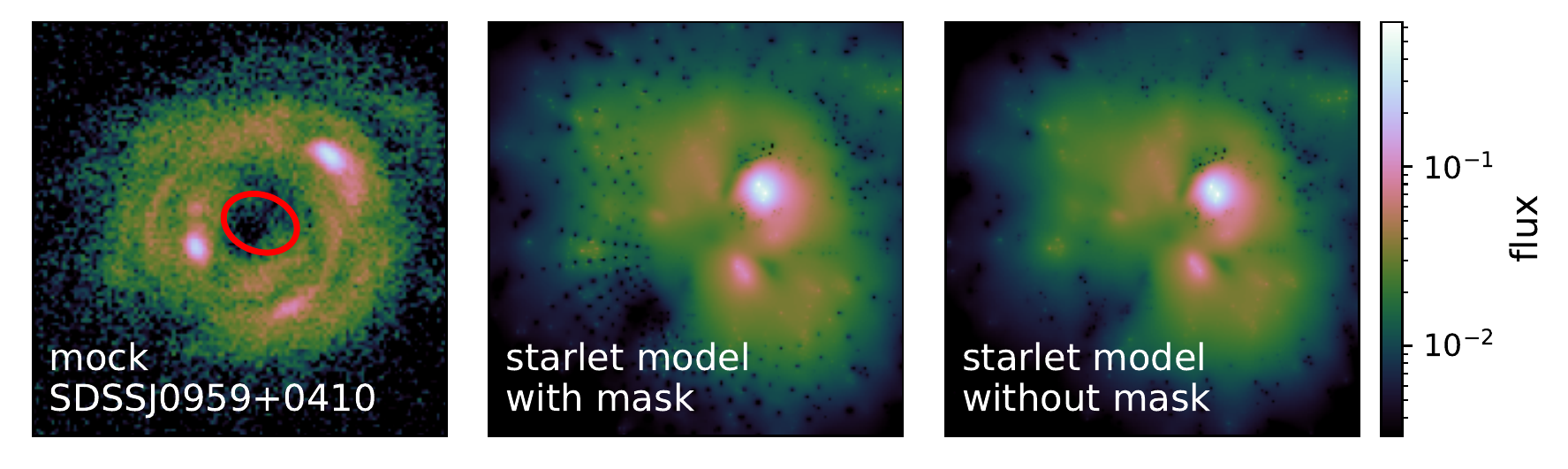}
    \label{app:fig:artifacts_origin}
    \caption{Origin of the artifacts seen in the starlet model of the SLACS system \zeroneufcinquanteneuf. \textbf{Left}: mock imaging data, with mask region covering the real lens light indicated in red. \textbf{Middle}: starlet reconstruction obtained with a mask on the central region (same as in Fig.~\ref{fig:slacs_fixed_mass}), where artifacts from inaccurate lens light modelling are clearly visible. \textbf{Right}: starlet reconstruction without masking the central region, which suppresses the artifacts. \href{https://github.com/aymgal/SLITronomy-papers/blob/master/paper_I/SLACS_source_reconstruction.ipynb}{\faGithub}}
\end{figure}

In Sect.~\ref{fig:slacs_fixed_mass}, we briefly discuss reasons for the artifacts visible in the \zeroneufcinquanteneuf source reconstruction with starlets. These artifacts are in part due to the pixelated approximation inherent to pixel-based modelling techniques, but mostly due to the complexity of the system and the difficulty of correctly modelling both the source and lens light distributions. To illustrate this, we create a realistic mock of the system \zeroneufcinquanteneuf by simulating lensed arcs based on our starlets source reconstruction, after downsampling it to the data resolution for removing the artifacts. We take the same PSF, exposure time, and background noise levels as in the real data. By construction, this also ensures no contamination by the lens light. We then perform a starlet reconstruction from the mock, with and without the central region at the location of the lens. We show both reconstructions side by side in Fig.
~\ref{app:fig:artifacts_origin}. We observe no artifacts in the reconstruction when not masking the central region, which confirms that these artifacts are primarily due to limitations of the analytical lens light modelling, followed by an inaccurate masking of the residual light.

\section{Effect of the number of starlet scales \label{app:effect_Js}}

\begin{figure}
    \centering
    \includegraphics[width=0.8\linewidth]{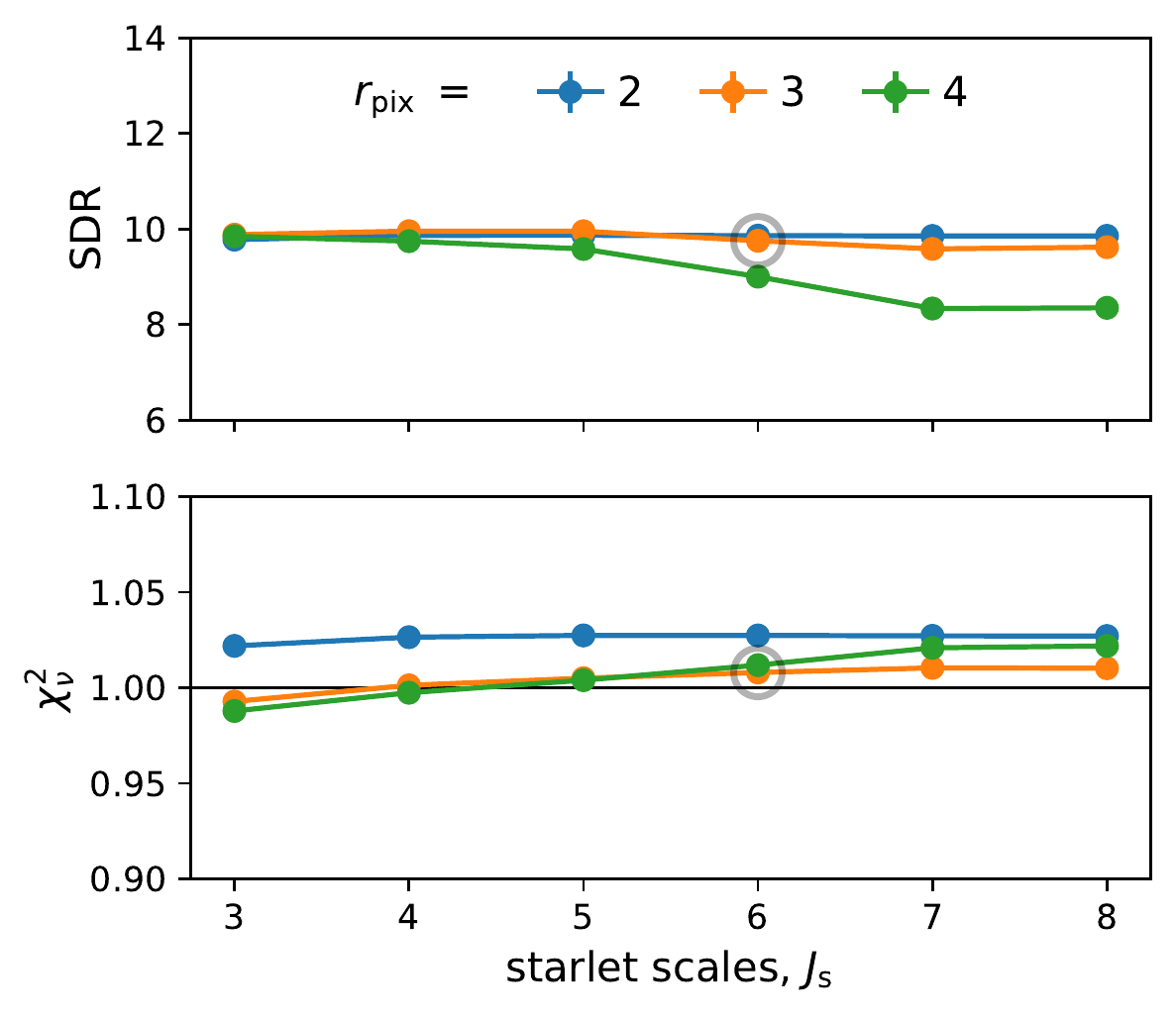}
    \caption{Effect of the choice of starlet decomposition scales, \Jscales{\source}, on the modelling quality, assessed by the two metrics introduced in Sect.~\ref{sssec:hst_starlets} The value $\Jscales{\source}=6$ chosen throughout this work is indicated by the gray circle. \href{https://github.com/aymgal/SLITronomy-papers/blob/master/paper_I/mock_source_reconstruction.ipynb}{\faGithub}}
    \label{app:fig:starlet_Js}
\end{figure}

In the implementation of our source reconstruction algorithm, which represents the surface brightness on a pixelated grid, we are subject to the choice of the source pixel size. In this work we introduce \pixsizeratio as the ratio between image and source pixel sizes. It is known that the choice of \pixsizeratio can impact the accuracy of lens model parameters (see Sect.~\ref{ssec:mass_sampling_mock_HST}). Here we assume a lens model that has been estimated accurately enough, for instance using the scheme shown on Fig~\ref{fig:starlets_refinement}.

The additional parameter specifically introduced by the starlet transform is the number of decomposition scales \Jscales{\source} and \Jscales{\lens} for source and lens light, respectively. We show in Fig~\ref{app:fig:starlet_Js} image plane ($\chi^2_\nu$) and source plane (SDR) metrics to assess the modelling quality for different \Jscales{\source} values. We take the source reconstruction of the single-component simulated HST data (top row of Fig.~\ref{fig:mock_HST}) as a test case. The number of decomposition scales has a small impact on the quality of the reconstruction, and we choose $\Jscales{\source}=6$ for reconstructions performed in this work. The number of decomposition scales can also simply be set to its maximal value, defined by the size of the region being regularised. For instance, if the source plane is defined on a grid of $n_{\rm pix}$ pixels on a side, the maximum number of scales allowed by the wavelet transform is $J_{\source,\rm max}=\lfloor\log_2{n_{\rm pix}}\rfloor$.

The choice of number of decomposition scale also affects the computation time considerably, as we show in Appendix \ref{app:benchmarks} below.

\section{\slitro benchmarks \label{app:benchmarks}}

\begin{figure*}
    \centering
    \includegraphics[width=\linewidth]{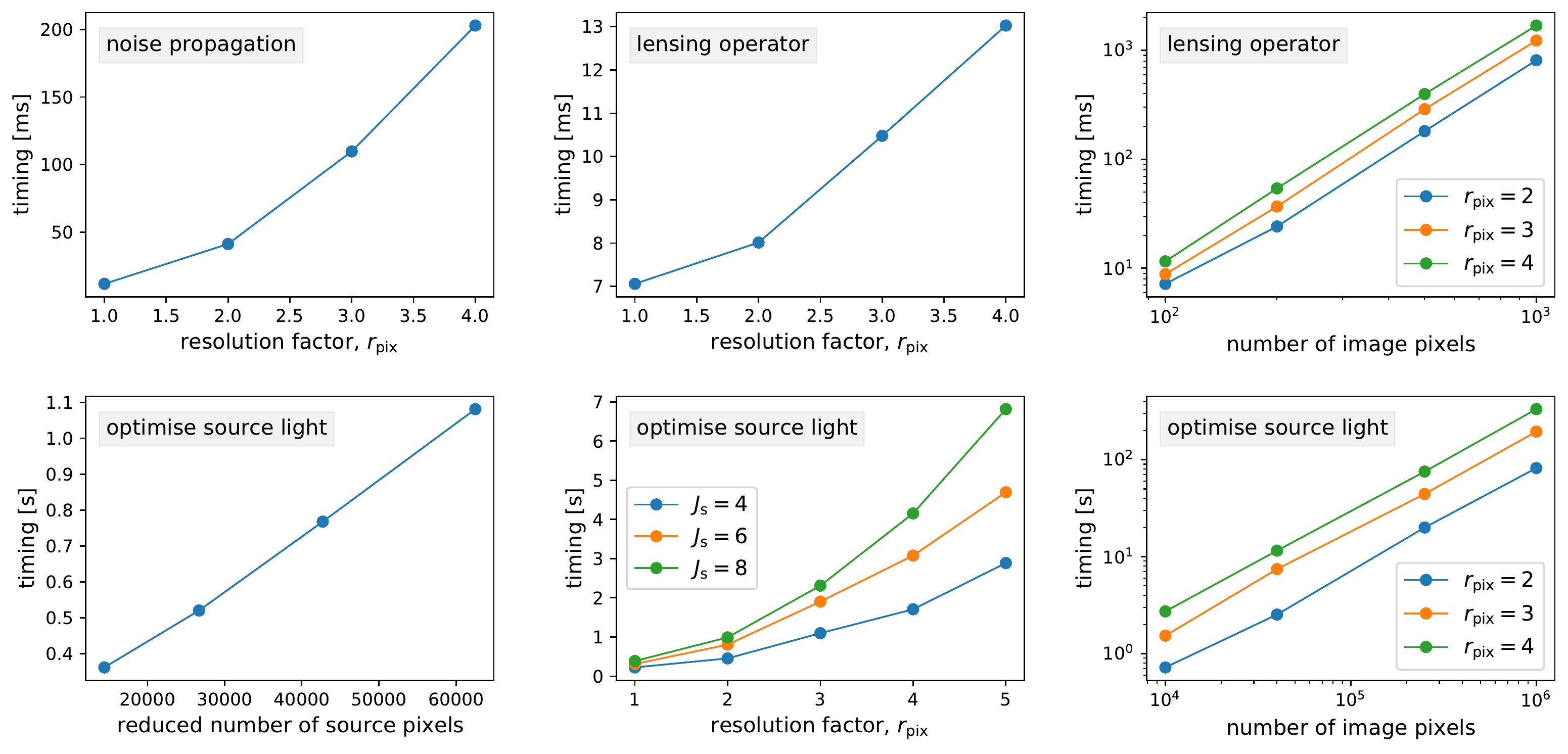}
    \caption{Runtimes for various settings of the sparse source solver of \slitro. \textbf{Top left}: propagation of the noise to transformed starlet space, for different source to image resolution \pixsizeratio. \textbf{Top center}: computation of the lensing operator \lensingop for different \pixsizeratio. \textbf{Top right}: computation of \lensingop for total number of data pixels. \textbf{Bottom left}: optimisation of the source light \source for different choices of the minimum allowed total number of source pixels. \textbf{Bottom center}: optimisation of \source for different \pixsizeratio and number of source starlet scales \Jscales{\source}. \textbf{Bottom right:} optimisation of \source for varying total number of data pixels. We refer to the text for more details. \href{https://github.com/aymgal/SLITronomy-papers/blob/master/paper_I/benchmarks.ipynb}{\faGithub}}
    \label{app:fig:runtimes}
\end{figure*}

In the current implementation of \slitro, lens model parameters can be optimised through different engines supported by \lenstro. This requires solving the sparse reconstruction problem for a given set of lens model parameters, computing the likelihood based on comparing the model with the imaging data, finally proposing a new set of lens model parameters, and so on. The total computation time required to find the best-fit lens model obviously depends heavily on the amount time necessary to solve Eq.~\ref{eq:slit_model}, as it will be solved at each iteration. To have a rough approximation of the computation time necessary to solve the sparse source reconstruction (\source in Eq.~\ref{eq:slit_model}), we give in Fig.~\ref{app:fig:runtimes} computation times for various settings of the algorithm and data sizes.

When not explicitly mentioned in the panels, the baseline choice is similar to our HST reconstructions in Sect.~\ref{sec:hst_data}: the simulated image is $100\times100$ pixels, with an image to source resolution $\pixsizeratio=3$ and starlet scales $\Jscales{\source}=6$.

In the top row we show timings for a subset of steps required for our algorithm to start, hence they are computed only once for a given lens model. They are shown in units of milliseconds. The top left panel shows the computation of the noise per pixel, propagated from image to source plane, including PSF effects and transformed in starlet space. This allows for a proper setting of the regularisation strength (see Sect.~\ref{sec:slit} for details). The top central and right panels show the computation of the lensing operator \lensingop for various numbers of image pixels and their corresponding number of source pixels through the choice of \pixsizeratio.

In the bottom row we consider the total time required to optimise the source light given a lens model. The first panel shows the runtime as a function of the number of source pixels, which can be automatically set to a minimal value, when for instance using a mask in the image plane (e.g. in Fig.~\ref{fig:source_recon_mock_ELT}). The masked region, once mapped back to source plane, defines a much smaller angular region on which the pixelated source light is defined, hence reducing the computation time considerably for both the lensing operator and subsequent mapping operations. The two following panels show the influence of the resolution factor \pixsizeratio, the number of image pixels, and the number of decomposition scales \Jscales{\source} on the computation time.

\end{appendix}

\typeout{get arXiv to do 4 passes: Label(s) may have changed. Rerun}

\end{document}